\documentclass{wiley-article}
\usepackage{ifthen}

\ifthenelse{\isundefined{\SUBMISSION}}
{\twocolumn}
{\onecolumn}

\usepackage{siunitx}

\usepackage{color}
\usepackage{amsmath}
\usepackage{amssymb}
\usepackage{graphicx}
\usepackage{placeins}
\usepackage{enumitem}
\usepackage[english]{babel}

\usepackage{tabto}
\usepackage{tikz}
\usetikzlibrary{shapes,arrows}
\usetikzlibrary[pgfplots.groupplots]
\usetikzlibrary{calc}
\usepackage{pgfplots}
\usetikzlibrary[pgfplots.groupplots]
\usepgfplotslibrary{fillbetween}
\usepackage{filecontents}
\usetikzlibrary{external}
\newlength{\figWidth}
\ifthenelse{\isundefined{\SUBMISSION}}
{\setlength{\figWidth}{\columnwidth}}
{\setlength{\figWidth}{0.6\columnwidth}}

\usepackage[pdftex,bookmarksnumbered,bookmarks=true,plainpages=false]{hyperref}
\hypersetup{colorlinks=true}
\hypersetup{linkcolor=blue}
\hypersetup{citecolor=blue}
\hypersetup{urlcolor=blue}

\usepackage{textcomp}
\usepackage{tabularx,colortbl}
\usepackage{todonotes}

\usepackage{setspace}

\ifthenelse{\isundefined{\SUBMISSION}}
{\newcommand{\zweizeilig}{}}
{\newcommand{\zweizeilig}{\doublespacing}}

\newcommand{\Giani}{s}
\newcommand{\Thiele}{Thiele modulus}

\papertype{Original Article}
\paperfield{Journal Section}

\title{Simulation of catalytic reactions in open-cell foam structures}

\author[1]{Sebastian M\"uhlbauer}
\author[1]{Severin Strobl}
\author[2\authfn{1}]{Matthew Coleman}
\author[1]{Thorsten P\"oschel}

\affil[1]{Institute for Multiscale Simulation, Friedrich-Alexander-Universit\"at Erlangen-N\"urnberg, Cauerstra{\ss}e 3, 91058 Erlangen, Germany} 
\affil[2]{Department of Chemical and Biomolecular Engineering, University of Notre Dame, 182 Fitzpatrick Hall, Notre Dame, IN 46556-5637} 

\corraddress{Sebastian M\"uhlbauer, Institute for Multiscale Simulation, Friedrich-Alexander-Universit\"at Erlangen-N\"urnberg, Cauerstra{\ss}e 3, 91058 Erlangen, Germany}
\corremail{sebastian.muehlbauer@fau.de }

\presentadd[\authfn{1}]{Institute for Multiscale Simulation, Friedrich-Alexander-Universit\"at Erlangen-N\"urnberg, Cauerstra{\ss}e 3, 91058 Erlangen, Germany}

\fundinginfo{DFG, SFB, ZISC, FPN, Haus Grant/Award Number: 123456}

\runningauthor{Sebastian M\"uhlbauer et al.}
\date{\today}
\paperreceived{\today}
\begin{document}

\begin{frontmatter}
\maketitle

\begin{abstract}
  We describe a technique for particle-based simulations of  heterogeneous catalysis in open-cell foam structures, which is based on isotropic Stochastic Rotation Dynamics (iSRD) together with Constructive Solid Geometry (CSG). The approach is validated by means of experimental results for the low temperature water-gas shift reaction in an open-cell foam structure modeled as inverse sphere packing.
  Considering the relation between Sherwood and Reynolds number, we find two distinct regimes meeting approximately at the strut size Reynolds number 10.
  For typical parameters from the literature, we find that the catalyst density in the washcoat can be reduced considerably without a notable loss of conversion efficiency.
  %
  %
  We vary the porosity to determine optimum open-cell foam structures, which combine low flow resistance with high conversion efficiency and find large porosity values to be favorable not only in the mass transfer limited regime but also in the intermediate regime.
  %
\end{abstract}
\end{frontmatter}
\zweizeilig 
\section{Introduction}

Open-cell foam structures are extremely promising substrates for heterogeneous catalysis~\cite{Twigg2002}. As a result of their high porosity, specific surface, and tortuosity, these structures provide excellent mass transport properties at moderate pressure drop~\cite{Lucci2015}. 
Therefore, the performance of open-cell foam structure for catalytic reactions was subject of several studies concerning both regular~\cite{Rickenbach2014, Rickenbach2015} and irregular structures~\cite{Lucci2014, Lucci2015}. In these studies, the foam structure was modeled as Kelvin cells, and continuum approaches such as finite volume methods were used to solve for the fields of concentration and flow velocity. Following this approach, the foam, which is of complicated geometrical shape, can be described by a correspondingly shaped mesh or by immersed boundaries~\cite{Peskin2002}. In both cases, the precision of the boundary description depends essentially on the spatial discretization of the simulation domain.
Smorygo et al.~\cite{Smorygo2011} followed a different approach and described open-cell foam structures by an inverse sphere packing, that is, the domain is given by the union of overlapping spheres and the characteristics of the foam are represented by the positions of the centers and the radii of the overlapping spheres. They could show that  inverse sphere packing models provide a good description for the specific surface and hydraulic permeability of foams. The great advantage of the inverse sphere packing model is that the simulation domain is defined analytically, thus the representation of the boundaries and consequently of the loci of the reactions is mathematically exact disregarding the resolution of the spatial discretization (lattice) of the domain. This way, the foam can be described as a hierarchy of spheres of significantly different size at reasonable computational effort.  

In the current paper, we employ isotropic Stochastic Rotation Dynamics (iSRD)~\cite{Muehlbauer2017} to simulate the flow of a reactive gas in a porous media. The foam structure is described using Constructive Solid Geometry (CSG)~\cite{Maentylae1987,STROBL2020}. Special care is required in regard to the boundary conditions, since pressure boundary conditions suffer from instabilities for particle-based fluid dynamics~\cite{Wang2004, Gatsonis2013}. On the other hand, periodic boundary conditions are obviously not adequate for the simulation of reactive flows. Therefore, here we use semi-periodic boundary conditions~\cite{Muehlbauer2018} which allow for a discontinuity of the concentration field, while the other flow fields, namely density, temperature, and flow velocity, are periodic. The flow is driven by an external acceleration.


For the verification of our simulation method we have chosen  the low temperature water-gas shift as prototype reaction due to its great industrial relevance. We assume the reaction to follow the Langmuir-Hinshelwood reaction mechanism~\cite{Fishtik2002, Ayastuy2005}. The foam structure serves as substrate and is assumed to be coated with CuO/ZnO/Al$_{2}$O$_{3}$ washcoat. We wish to point out that this study is not restricted to the mass transfer limited regime, but also the reaction rate limited regime is addressed. Therefore, it is eminently important to model the reaction mechanism in detail. The effective reaction rate in the washcoat layer is quantified by means of precomputed look-up tables for the effectiveness factor~\cite{Roberts1966}. Among other parameters, the effective reaction rate depends on the partial surface pressures of the reactants, which can be obtained from the collision fluxes on the surface. Hence, the relevant quantities are obtained directly at the catalytic boundary. 

In the following section, we describe our simulation method and the model for the chemical reaction. In Sec.~\ref{sec:validate}, we validate the individual parts of our method with experimental data taken from the literature and in Sec.~\ref{sec:simul-based-pred}, we study the properties of the low temperature water-gas shift reaction in porous foam for a range of parameters. Here, we vary both the active site density within the washcoat and the porosity of the unit foam cell. For the mass transfer controlled regime, Lucci et al. find that the dimensionless performance index increases monotonously with the porosity~\cite{Lucci2014}. We will reproduce this result for the low temperature water-gas shift and explore the validity of this relation towards the reaction rate limited regime.

\section{Numerical Model}

\subsection{Open-cell foam}
\label{sec:foam}

Smorygo et al. have show that inverse sphere packings are an excellent model for open-cell porous foam structures~\cite{Smorygo2011}, therefore, we use constructive solid geometry (CSG)~\cite{Maentylae1987,STROBL2020} to model the geometrical shape of the foam. This shape is then used as boundary of the simulation for the particle based fluid dynamics described in Sec. \ref{sec:gas_dynamics}.

\begin{figure}[!htb]
 \centering
 \tikzsetnextfilename{BasisCell}
 \tikzset{external/export next=false}
 	\begin{tikzpicture}
 	\node at (0,0){
 		\includegraphics[width=\figWidth]{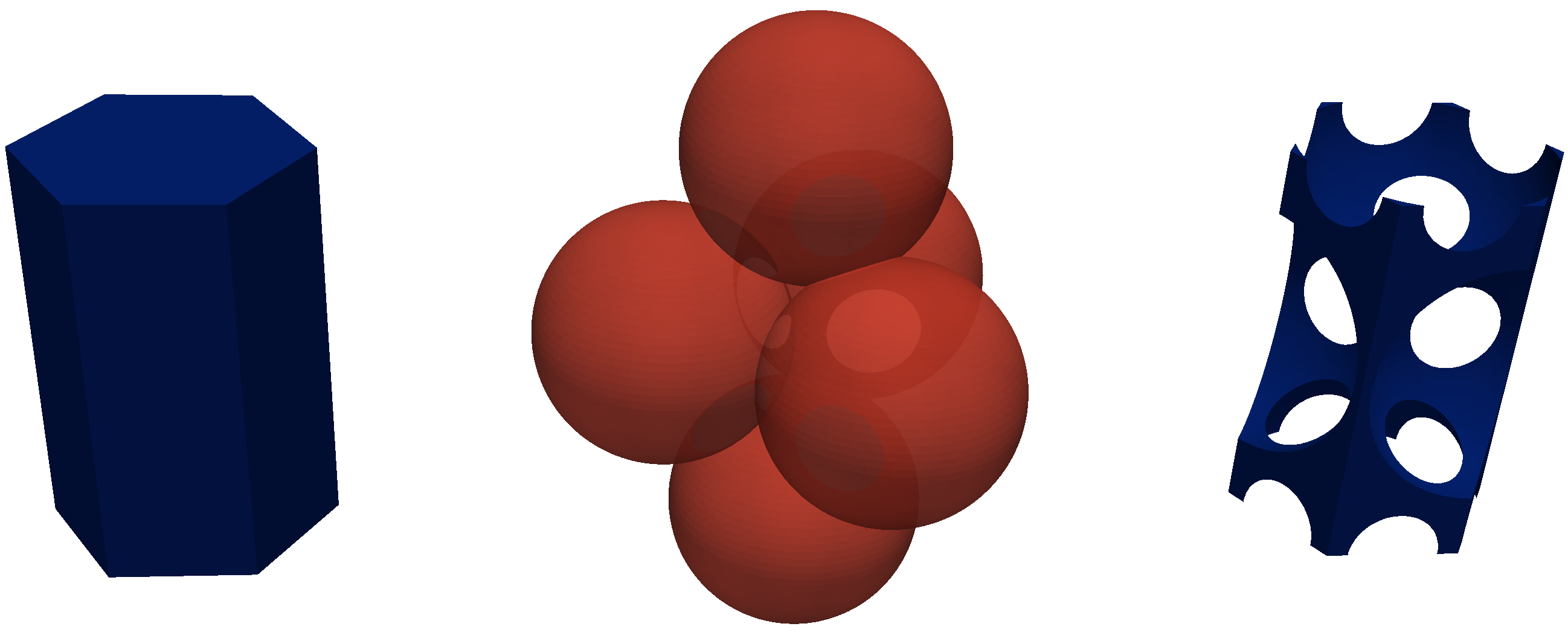}
 	};
 	%
 	\node (-) at (-0.185\figWidth,0\figWidth) {$ \setminus $};
 	\node (=) at (0.185\figWidth,0\figWidth) {$ = $};
 	\end{tikzpicture}
 	\caption{\zweizeilig The open-cell foam structure is modeled as an inverse sphere packing, resulting from the geometric subtraction of a sphere packing from a hexagonal prism. The porosity depends on the ratio between the sphere diameter and the width of the prism.}
 	\label{fig:CSG}
\end{figure}

Figure \ref{fig:CSG} illustrates the construction of our unit cell: the foam geometry results from the geometric subtraction of the sphere packing from the hexagonal prism. This approach allows us to fully parametrize the geometry by the prism width, $w$, and the sphere diameter, $d$. The porosity of the foam is controlled by the ratio $k \equiv w/d$.
Thus, the prism width, $w$, represents the pore size of the foam.


The CSG concept works particularly well for particle based simulation methods, where the boundary conditions can be enforced directly on every particle hitting the wall. Moreover, this approach allows us to evaluate the partial pressure corresponding to each individual particle species exactly at the surface via the collision fluxes, that is, simply by bookkeeping over the particles having hit the wall.

The numerically robust implementation of CSG boundaries to represent complex shapes is nontrivial; details can be found in~\cite{STROBL2020,Strobl2017}.

\subsection{Gas flow}
\label{sec:gas_dynamics}

\subsubsection{Fluid model}
\label{sec:fluid-model}

The gas flow is simulated by means of a variant of Stochastic Rotation Dynamics~\cite{Malevanets1999, Gompper2009}. In SRD, the fluid is modeled by point-like quasi-particles which do not directly correspond to actual particles, but represent a discretization of the phase space. Each time step comprises one streaming step, propagating the particles according to their current velocity, and one collision step, allowing the particles to exchange momentum. SRD has been shown to be a reliable to model for fluid flow through complex geometries~\cite{Lamura2001, DeAngelis2012}. Here, we use isotropic Stochastic Rotation Dynamics (iSRD)~\cite{Muehlbauer2017}, a variant of standard SRD. While the streaming step is identical in SRD and iSRD, for the collision step, in iSRD the particles are grouped into randomly distributed spheres, and not into the cubic cells of a Cartesian simulation grid as in SRD. Hence, iSRD unlike SRD does not suffer from anisotropy at complex shaped domain boundaries~\cite{Muehlbauer2017}.

\subsubsection{Boundaries}
\label{sec:boundaries}

\emph{Inlet and outlet.} For the simulation of the gas flow we assume semi-periodic boundary conditions~\cite{Muehlbauer2018} in flow direction. This means, we apply periodic boundary conditions at the inlet and outlet for the fields of density, velocity, temperature, and pressure while for the field of the composition of the  simulated gas mixture, we allow for a discontinuity at the (otherwise) periodic boundary. This type of boundary model was designed particularly to describe reactive gas flows. It assures temporally constant reactant and product concentrations at the inlet. It has been shown that this boundary condition is suitable to mimic reactive flows with pressure boundary conditions at much lower computational effort~\cite{Muehlbauer2018}.

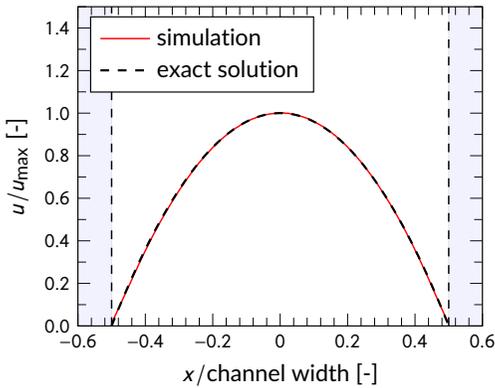
\begin{figure}[!htb]
 \centering
 \tikzsetnextfilename{Poiseuille}
 \resizebox{\figWidth}{!}{%
 \begin{tikzpicture}
   \begin{axis}
   [
     axis line style={-},
     xlabel={$ x / \text{channel width} $ [-]},
     ylabel={$ u / u_{\text{max}} $ [-]},
     xlabel near ticks,
     ylabel near ticks,
     xmin=-0.6, xmax=0.6,
     ymin=0, ymax=1.5,
     tick style={color=black},
     tick label style={font=\footnotesize},
     y tick label style={
       /pgf/number format/.cd,
       fixed,
       fixed zerofill,
       precision=1,
       /tikz/.cd },
     xtick={-0.6,-0.4,-0.2,0,0.2,0.4,0.6},
     ytick={0,0.2,...,1.6},
     minor x tick num={4},
     minor y tick num={1},
     width=0.95\figWidth,
     height=0.80\figWidth,
     legend cell align={left},
     legend pos=north west,
   ]
   \draw[thin,fill=blue!50,opacity=0.1] (axis cs:\pgfkeysvalueof{/pgfplots/xmin},\pgfkeysvalueof{/pgfplots/ymin}) -- (axis cs:-0.5,\pgfkeysvalueof{/pgfplots/ymin}) -- (axis cs:-0.5,\pgfkeysvalueof{/pgfplots/ymax}) -- (axis cs:\pgfkeysvalueof{/pgfplots/xmin},\pgfkeysvalueof{/pgfplots/ymax}) -- cycle;
   \draw[thin,fill=blue!50,opacity=0.1] (axis cs:0.5,\pgfkeysvalueof{/pgfplots/ymin}) -- (axis cs:\pgfkeysvalueof{/pgfplots/xmax},\pgfkeysvalueof{/pgfplots/ymin}) -- (axis cs:\pgfkeysvalueof{/pgfplots/xmax},\pgfkeysvalueof{/pgfplots/ymax}) -- (axis cs:0.5,\pgfkeysvalueof{/pgfplots/ymax}) -- cycle;
   \addplot[solid,line width=0.5pt,no markers,red] table[x expr=((\thisrowno{0} - 2.43e-4) / 4.344e-4), y expr=(\thisrowno{3} / 146.5), col sep=space] {./figs/Data/ObstacleBoundary/velocity-profile.dat};
   \addlegendentry{simulation};
   \addplot[dashed,line width=0.75pt,no markers,black,domain=-0.5:0.5,samples=200] {1 - 4*x^2};
   \addlegendentry{exact solution};
   \addplot[dashed,line width=0.5pt,samples=10, smooth,domain=0:1,black, name path=three] coordinates {(-0.5,\pgfkeysvalueof{/pgfplots/ymin})(-0.5,\pgfkeysvalueof{/pgfplots/ymax})};
   \addplot[dashed,line width=0.5pt,samples=10, smooth,domain=0:1,black, name path=three] coordinates {(0.5,\pgfkeysvalueof{/pgfplots/ymin})(0.5,\pgfkeysvalueof{/pgfplots/ymax})};
   \end{axis}
 \end{tikzpicture}
 }
 \caption{\zweizeilig Test of the described boundary conditions for the case of plane Poiseuille flow. Obviously, the desired no-slip boundary conditions at the solid wall are obtained and at the same time the analytical result is reproduced up to very good precision.}
 \label{fig:Poiseuille}
 \end{figure}

\emph{Solid wall.} In order to simulate no-slip conditions at the solid walls formed by the foam described in Sec. \ref{sec:foam}, the velocity of each fluid particle hitting the wall is inverted. This is, however, not sufficient to assure no-slip conditions if the collision bins (in SRD) or the collision spheres (in iSRD) overlap with the domain boundary~\cite{Lamura2001}. Therefore, we generate ghost particles within the solid walls. Their velocities are randomly chosen from a Maxwell-Boltzmann distribution. The number density of the ghost particles is chosen identical to the number density of particles in the fluid domain~\cite{Bolintineanu2012}. To demonstrate that this procedure delivers indeed reliable results, in Fig. \ref{fig:Poiseuille} we show the numerically obtained flow velocity profile for plane Poiseuille flow. 
The gray shaded areas indicate the regions populated by ghost particles. We see that the simulation data agrees up to a high precision with the analytical solution for plane Poiseuille flow. 

\subsubsection{Reaction model}
\label{sec:reaction-model}

For demonstration of the model and the algorithm, we have chosen the low temperature water-gas shift reaction for its great importance in chemical engineering, that is, the production of carbon dioxide and hydrogen from carbon monoxide and water steam,
\begin{equation}
\label{eq:WGS}
  \text{CO} + \text{H}_{2}\text{O} \rightarrow \text{CO}_{2} + \text{H}_{2} \, .
\end{equation}
The back reaction is not taken into account. This reaction proceeds along the \emph{Langmuir-Hinshelwood} reaction path~\cite{Ayastuy2005} which is one of the most important prototype reaction paths for heterogeneous catalysis. Therefore, the results and conclusions attained for this example are applicable similarly also to other catalytic reactions. In industrial applications the carrier structure is often coated by a porous washcoat layer. This washcoat provides a large effective surface containing catalytic sites needed for the water-gas shift reaction to take place. We assume the foam structure to be coated evenly and denote the washcoat layer thickness by $L$. Since, the reactant concentrations decrease with increasing depth in the washcoat, the local reaction rate in the washcoat, defined as the number of reactions per unit time and washcoat volume, $ \text{d}n_r/\text{d}t$, varies as well. Note that due to the Langmuir-Hinshelwood mechanism, the reaction rate is not necessarily a monotonous function of the reactant concentrations~\cite{Roberts1966}.

The gas flow through the mesoporous foam structure is simulated as described in Sec. \ref{sec:fluid-model}. Albeit both foam and washcoat have a similar geometric structure, it is not feasible to model the open-cell foam and the washcoat simultaneously, since the associated processes occur on very different length and time scales. Hence, in the simulation the reaction-diffusion process related to the washcoat is decoupled from the particle simulation by introducing an effective model for the reactions. Again, we neglect the back reaction and obtain for the average reaction rate in the washcoat layer
\begin{equation}
  \label{eq:reaction_rate}
  \left< \frac{\text{d}n_r}{\text{d}t}  \right>  = \zeta\,R
\end{equation}
with
\begin{equation}
  \label{eq:R}
R=  \frac{ k  P_\text{CO} P_{\text{H}_2\text{O}} }{ \left(1\! +\! K_\text{CO} P_\text{CO}\! +\! K_{\text{H}_2\text{O}} P_{\text{H}_2\text{O}}\! +\! K_{\text{CO}_2} P_{\text{CO}_2}\! +\! K_{\text{H}_2} P_{\text{H}_2}\right)^2 } 
\end{equation}
where $k$ is the reaction rate constant and $K_i$ and $ P_i $ with $i\in\left\{\text{CO},\text{H}_2\text{O},\text{CO}_2, \text{H}_2\right\}$ are the adsorption constant and the partial pressure of the component $i$ at the outer surface of the washcoat, respectively. Thus, the average reaction rate, Eq. \eqref{eq:reaction_rate}, of unit $\text{mol} / (\text{m}^3 \, \text{s})$ results from the local reaction rate at the outer surface of the washcoat, $R$, given by Eq. \eqref{eq:R},  and the effectiveness factor, $\zeta$.

We follow the procedure proposed in~\cite{Roberts1966} to compute the effectiveness factor for the relevant range of the partial reactant pressures and the model parameters, which vary with temperature. Note that for the computation of the effectiveness factor for a given parameter set, the one-dimensional reaction-diffusion equation for Langmuir-Hinshelwood kinetics in the washcoat has to be solved~\cite{Roberts1966}, while the reaction rate directly at the surface can be evaluated very easily.

The domain boundary described by the inverse sphere packing corresponds to the outer surface of the washcoat. The catalytic boundary is discretized into individual boundary cells. The partial surface pressures are computed from the given wall temperature and the temporally averaged collision fluxes, $Z_i$, with $i\in\left\{\text{CO},\text{H}_2\text{O},\text{CO}_2, \text{H}_2\right\}$. Subsequently, the effectiveness factor is read from the precomputed table, and the average reaction rate follows from Eq. \eqref{eq:reaction_rate}. To transform
the simulated quasi-particles according to the chemical reaction, we need the reaction probability for each reactant species, which is obtained via
\begin{equation}
  \label{eq:reaction_prob}
  p_i  = \left< \frac{\text{d}n_r}{\text{d}t}\right> \, L \, A\, \frac{ \nu_i }{ Z_i } \, , \ \ \ \text{for} \ i \in \left\{ \text{CO}, \, \text{H}_2\text{O} \right\} \, ,
\end{equation}
where $L$ and $A$ denote the washcoat thickness and the surface area in the concerned boundary cell, respectively. Thus, whenever a particle of type $ i $ hits the catalytic boundary, this particles undergoes a reaction with probability $ p_i $, which has to be computed beforehand in the corresponding boundary cell. For the reaction considered here, the stoichiometric coefficient, $\nu_i=1$, for both reactants. This approach, on average, recovers the expected number of conversions for each reactant species.


\paragraph{Washcoat parameters}
\label{sec:wcParams}

For the current study, we refer to Ayastuy et al.~\cite{Ayastuy2005}. who assessed various approaches to model the low temperature water-gas shift reaction. In their experiments, the catalyst is composed of $24.9 \, \% \ \text{CuO}$, $43.7 \, \% \ \text{ZnO}$ and $31.4 \, \% \ \text{Al}_2\text{O}_3$. Having a pore volume of $0.29 \, \text{cm}^3$ per gram catalyst, this yields an effective catalyst density of $2.06 \, \text{g} / \text{cm}^3$ including the pores. The characteristic pore diameter in the catalyst is in the range of $10\dots 20 \, \text{nm} $. The most accurate model considered in~\cite{Ayastuy2005} is based on the Langmuir-Hinshelwood reaction mechanism. For the temperature, $ T = 453 \, \text{K} $, the kinetic constant for this model is $ k = 1.217 \times 10^{4} \, \text{mol} / (\text{m}^3 \, \text{s} \, \text{Pa}^2) $, where the volume refers to catalyst volume. The kinetic constant is assumed proportional to the active site density. Thus, lower active site density is achieved by reducing $k$. The adsorption constants referring to the different species are:
\begin{equation}
  \label{eq:1}
  \begin{split}
 K_\text{CO} &= 8.562 \times 10^{-3} \,\text{Pa}^{-1} \\
K_{\text{H}_2\text{O}} &= 8.898 \times 10^{-4} \, \text{Pa}^{-1}\\
K_{\text{CO}_2} &= 4.581 \times 10^{-4} \, \text{Pa}^{-1}\\
K_{\text{H}_2} &= 9.594 \times 10^{-4} \, \text{Pa}^{-1}\,.    
  \end{split}
\end{equation}
The geometric washcoat properties are taken from sample A15 examined by Novak et al. in~\cite{Novak2012}. Here the washcoat thickness is $ L = 15 \, \text{\textmu m} $. Below we will also need the macro- and mesoporosity given as $\varepsilon_\text{macro} = 0.26 $ and $ \varepsilon_\text{meso} = 0.43 $, respectively. The pore diameter, required to compute the Knudsen diffusion coefficients in the mesopores, is $13\,\text{nm} $. The effective diffusion coefficients in the washcoat are estimated based on what Novak et al.~call the \textquotedblleft standard model\textquotedblright~\cite{Novak2012}:
\begin{equation}
  \label{eq:2}
  \begin{split}
D_\text{CO}^\text{wc} & = 5.068 \times 10^{-6} \, \text{m}^2 / \text{s}\\
D_{\text{H}_2\text{O}}^\text{wc} &= 7.342 \times 10^{-6} \, \text{m}^2 / \text{s}\\
D_{\text{CO}_2}^\text{wc} &= 4.425 \times 10^{-6} \, \text{m}^2 / \text{s}\\
D_{\text{H}_2}^\text{wc} &= 1.553 \times 10^{-5} \, \text{m}^2 / \text{s}\,.
  \end{split}
\end{equation}
%

\section{Validation}
\label{sec:validate}

\subsection{Pressure drop -- Hagen number}
\label{sec:pressure-drop-hagen}

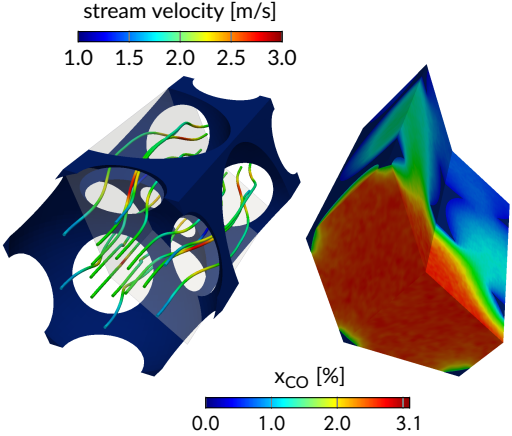
\begin{figure}[!htb]
\centering
\tikzsetnextfilename{Foam}
\tikzset{external/export next=false}
	\begin{tikzpicture}
	\node at (0,0){
		\includegraphics[width=\figWidth]{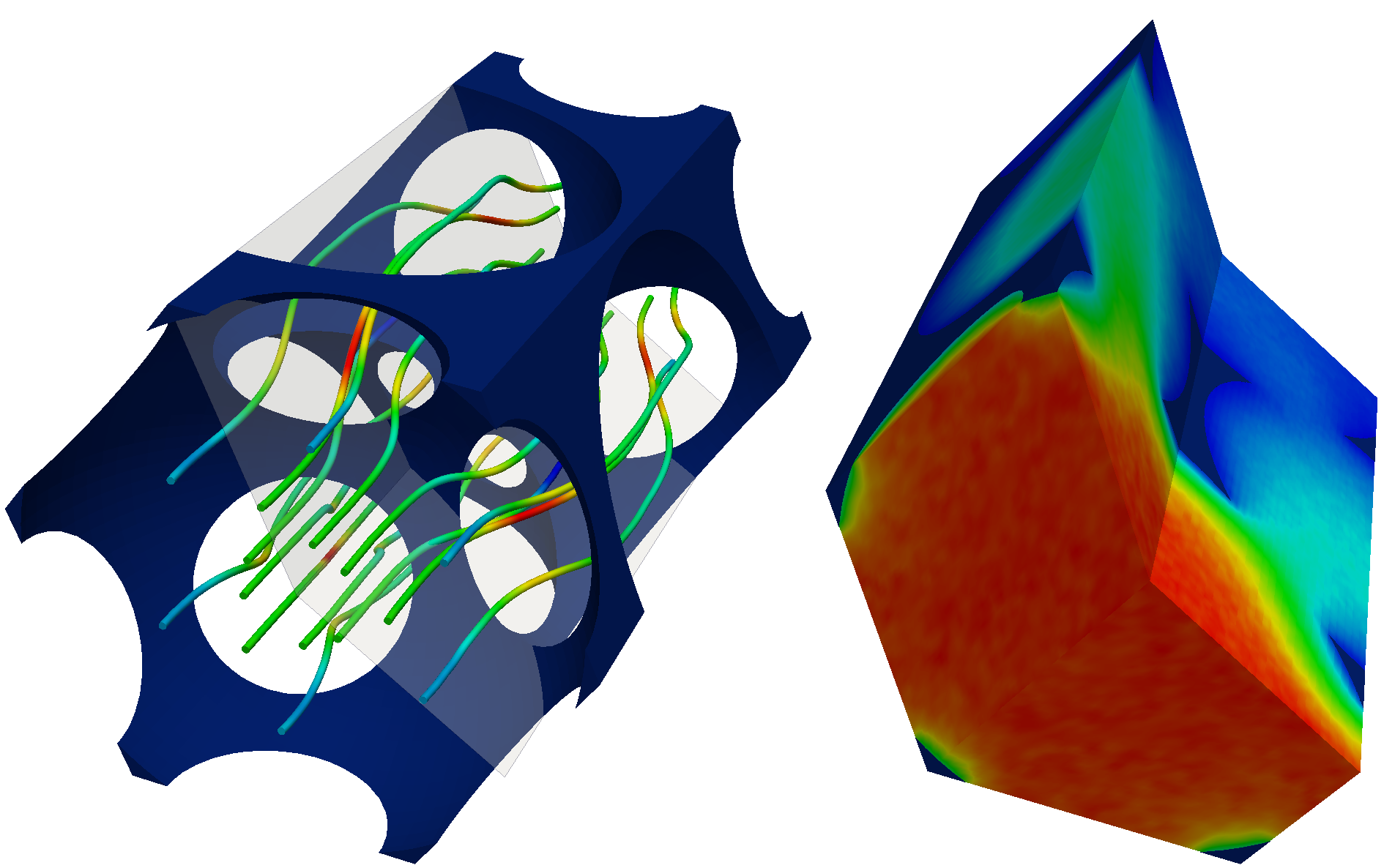}
	};
	\begin{axis}
	[
		at={(-0.35\figWidth,0.41\figWidth)},
		hide axis,
		scale only axis,
		height=0pt,
		width=0pt,
		colormap/jet,
		colorbar horizontal,
		point meta min=16,
		point meta max=48,
		colorbar style={
			width=0.4\figWidth,
			height=0.02\figWidth,
			xtick={16,24,...,48},
			xticklabels={1.0,1.5,2.0,2.5,3.0}
		}
	]
		\addplot [draw=none] coordinates {(0,0)};
	\end{axis}
	\begin{axis}
	[
		at={(-0.10\figWidth,-0.31\figWidth)},
		hide axis,
		scale only axis,
		height=0pt,
		width=0pt,
		colormap/jet,
		colorbar horizontal,
		point meta min=0.0,
		point meta max=3.1,
		colorbar style={
			width=0.4\figWidth,
			height=0.02\figWidth,
			xtick={0,1,2,3,3.1},
			xticklabels={$0.0$,$1.0$,$2.0$,,$3.1$}
		}
	]
		\addplot [draw=none] coordinates {(0,0)};
	\end{axis}
	\node (streamvelocity) at (-0.15\figWidth,0.405\figWidth) {stream velocity [m/s]};
	\node (concentration) at (0.105\figWidth,-0.315\figWidth) {$\text{x}_\text{CO}$ [$\%$]};
	\end{tikzpicture}
	\caption{\zweizeilig Typical simulation result. The figure shows the fields of flow velocity (left) and concentration (right) in a periodic unit foam cell, represented by color and streamlines, respectively. In this example the porosity is $\varepsilon = 0.902$. The remaining parameters are taken from Tab. \ref{tab:setup} explained in detail in Sec. \ref{sec:simul-setup}.}
	\label{fig:Foam}
\end{figure}

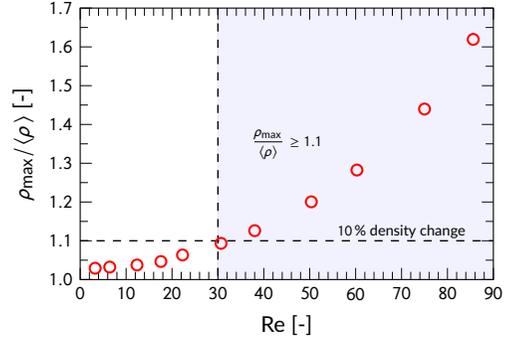
\begin{figure}[!htb]
\centering
\tikzsetnextfilename{Compression}
\resizebox{\figWidth}{!}{%
\begin{tikzpicture}
  \begin{axis}
  [
    axis line style={-},
    xlabel={Re [-]},
    ylabel={$ {\rho_{\text{max}}}/{\langle \rho \rangle} $ [-]},
    xlabel near ticks,
    ylabel near ticks,
    xmin=0, xmax=90,
    ymin=1, ymax=1.7,
    tick style={color=black},
    tick label style={font=\footnotesize},
    y tick label style={
      /pgf/number format/.cd,
      fixed,
      fixed zerofill,
      precision=1,
      /tikz/.cd },
    xtick={0,10,...,90},
    ytick={1,1.1,...,1.8},
    minor x tick num={4},
    minor y tick num={1},
    width=0.95\figWidth,
    height=0.70\figWidth,
    legend cell align={left},
    legend pos=north west,
  ]
  \draw[thin,fill=blue!50,opacity=0.1] (axis cs:30,\pgfkeysvalueof{/pgfplots/ymin}) -- (axis cs:90,\pgfkeysvalueof{/pgfplots/ymin}) -- (axis cs:90,\pgfkeysvalueof{/pgfplots/ymax}) -- (axis cs:30,\pgfkeysvalueof{/pgfplots/ymax}) -- cycle;
  \addplot[only marks,line width=0.75pt,mark=o,mark size=2.0pt,red] table[x expr=(\thisrowno{14}), y expr=(\thisrowno{20}), col sep=space] {figs/Data/Validation/k1100/results_FINAL.dat};
%
   \addplot[dashed,line width=0.5pt,no markers,black,domain=0:90,samples=200] {1.10};
%
      \addplot[dashed,line width=0.5pt,samples=10, smooth,domain=0:1,black, name path=three] coordinates {(30,0)(30,55)};
  \node[] at (axis cs: 70,1.125) {\scriptsize $10 \, \%$ density change};
  \node[] at (axis cs: 45,1.35) {\scriptsize $\dfrac{\rho_{\text{max}}}{\langle \rho \rangle} \geq 1.1 $};
  \end{axis}
\end{tikzpicture}
}
\caption{\zweizeilig Maximum compression within the simulation domain as a function of Reynolds number. The dashed line separates regions with $\rho_\text{max}/\left<\rho\right> < 10\%$ from regions with larger deviation. This separation corresponds to $\text{Re}=30$.}
\label{fig:compression}

\end{figure}

A typical simulation result is shown in Fig. \ref{fig:Foam} where the concentration field and the field of flow velocity are represented by color and stream lines, respectively.
From the flow fields obtained from the simulations we determine the superficial velocity, $U$, and the pore size Reynolds number,
\begin{equation}
  \label{eq:Rey}
  \text{Re} \equiv \dfrac{w U}{\nu} \, ,
\end{equation}
where $w$ is the prism width and $\nu$ denotes the kinematic viscosity. Further, we compute the dimensionless pressure drop represented by the Hagen number,
\begin{equation}
  \label{eq:Hg}
  \text{Hg} \equiv -\dfrac{w^3}{\left< \rho \right> \, \nu^2} \, \frac{\text{d}P}{\text{d}z} \, ,
\end{equation}
where $P$ is the pressure, $z$ is the coordinate along the flow direction, and $\langle \rho \rangle$ is the average mass density in the system. In mesoscopic simulations~\cite{Kikuchi2003, Gompper2009, Muehlbauer2017}, the Reynolds number can be increased either by increasing the spatial and temporal resolution (which leads to a decrease of viscosity) or by increasing the flow velocity. Here, we vary the superficial velocity, while the pore size and the resolution and, consequently also the kinematic viscosity remain invariant. For $\text{Re}\gtrsim 30$, this approach leads to density variations of more than $10 \, \%$, as shown in Fig. \ref{fig:compression}. 

Even for $\text{Re} \rightarrow 0$, the value of $ \rho_\text{max}/\left< \rho \right>$ would not approach the ideal value of $1.0$ due to fluctuations in density and velocity which are inherent to the statistical nature of iSRD and similar particle-based models of CFD. Thus, part of the density variation shown in Fig. \ref{fig:compression} must be attributed to statistical fluctuations of the system. To obtain the results shown in this manuscript, we average over $ 2 \times 10^4 $ time steps, after the steady state has been reached.

\begin{figure}[!htb]
\centering
\tikzsetnextfilename{HgRe}
 \resizebox{\figWidth}{!}{%
\begin{tikzpicture}
  \begin{axis}
  [
    axis line style={-},
    xlabel={Re [-]},
    ylabel={Hg [-]},
    xlabel near ticks,
    ylabel near ticks,
    xmin=0, xmax=90,
    ymin=0, ymax=2e4,
    tick style={color=black},
    tick label style={font=\footnotesize},
    y tick label style={
      /pgf/number format/.cd,
      fixed,
      fixed zerofill,
      precision=1,
      /tikz/.cd },
    xtick={0,10,...,90},
    minor x tick num={4},
    minor y tick num={4},
    width=0.95\figWidth,
    height=0.70\figWidth,
    legend cell align={left},
    legend pos=north west,
  ]
  \draw[thin,fill=blue!50,opacity=0.1] (axis cs:30,\pgfkeysvalueof{/pgfplots/ymin}) -- (axis cs:90,\pgfkeysvalueof{/pgfplots/ymin}) -- (axis cs:90,\pgfkeysvalueof{/pgfplots/ymax}) -- (axis cs:30,\pgfkeysvalueof{/pgfplots/ymax}) -- cycle;
  \addplot[only marks,line width=0.75pt,mark=o,mark size=2.0pt,red] table[x expr=(\thisrowno{14}), y expr=(\thisrowno{15}), col sep=space] {figs/Data/Validation/k1100/HagenNumberDukhan2006/pressure_drop_deviation.dat};
  \addlegendentry{simulation, $\varepsilon = 0.914$};
   \addplot[solid,line width=0.75pt,no markers,black] table[x expr=(\thisrowno{8}), y expr=(\thisrowno{9}), col sep=space] {figs/Data/Validation/k1100/HagenNumberDukhan2006/Dukhan2006_Figure2i_0.915.dat};
   \addlegendentry{experiment, $\varepsilon = 0.915$};
      \addplot[dashed,line width=0.5pt,samples=10, smooth,domain=0:1,black, name path=three] coordinates {(30,0)(30,2e4)};
  \node[] at (axis cs: 45,1.0e4) {\scriptsize $\dfrac{\rho_{\mathit{max}}}{\langle \rho \rangle} \geq 1.1 $};
  \end{axis}
\end{tikzpicture}
}
\caption{\zweizeilig Dimensionless pressure drop quantified by the Hagen number, Eq. \eqref{eq:Hg}, as a function of the Reynolds number. Symbols: simulation results; soid line: experimental data (sample $2$ in~\cite{Dukhan2006}, see Fig. 2(i) there). Dashed line: see caption to Fig. \ref{fig:compression}.}
\label{fig:HgRe}
\end{figure}
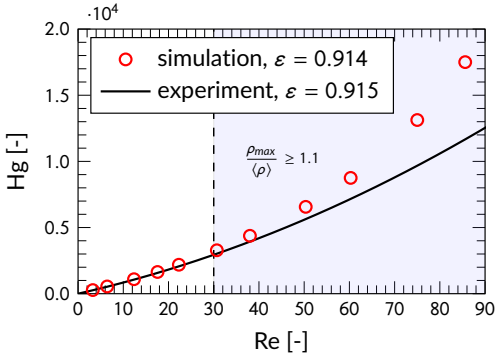

In Figure \ref{fig:HgRe},  we show the relation between the Hagen number, representing the pressure drop, and the Reynolds number, $\text{Hg}=\text{Hg}(\text{Re})$. 
For $\text{Re} \lesssim 30$, the deviations between simulation results and experimental data, taken from~\cite{Dukhan2006}, remain below $6 \, \%$. These deviations are due to the imperfect representation of the foam geometry by the inverse sphere packing. For $\text{Re} \gtrsim 30$, the deviations grow with increasing Reynolds number. This growth is associated with the compressibility effects, depicted in Fig. \ref{fig:compression}. These effects  play a minor role below $\text{Re} = 30$. We conclude that our model and simulation method reliably predicts the pressure drop for metal foam structures in the low Reynolds number regime.

\subsection{Mass transport -- Sherwood number}
\label{sec:mass-transp-sherw}

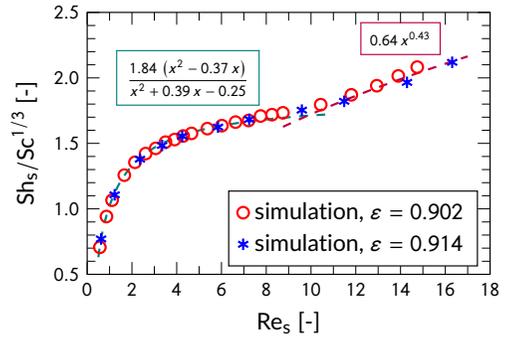
\begin{figure}[!htb]
\centering
\tikzsetnextfilename{ShRe}
 \resizebox{\figWidth}{!}{%
\begin{tikzpicture}
  \begin{axis}
  [
    axis line style={-},
    xlabel={$\text{Re}_\text{\Giani}$ [-]},
    ylabel={$\text{Sh}_\text{\Giani} / \text{Sc}^{1/3}$ [-]},
    xlabel near ticks,
    ylabel near ticks,
    xmin=0, xmax=18,
    ymin=0.5, ymax=2.5,
    y tick label style={
      /pgf/number format/.cd,
      fixed,
      fixed zerofill,
      precision=1,
      /tikz/.cd },
    tick style={color=black},
    tick label style={font=\footnotesize},
    xtick={0,2,...,18},
    ytick={0,0.5,...,3},
    minor x tick num={1},
    minor y tick num={4},
    width=0.95\figWidth,
    height=0.70\figWidth,
    legend cell align={left},
    legend pos=south east,
  ]
  \addplot[only marks,line width=0.75pt,mark=o,mark size=2.0pt,red] table[x expr=(\thisrowno{30}), y expr=(\thisrowno{32}), col sep=space] {figs/Data/Validation/k1090/results_FINAL.dat};
  \addlegendentry{simulation, $\varepsilon = 0.902$};
  \addplot[only marks,line width=0.75pt,mark=asterisk,mark size=2.0pt,blue] table[x expr=(\thisrowno{30}), y expr=(\thisrowno{32}), col sep=space] {figs/Data/Validation/k1100/results_FINAL.dat};
  \addlegendentry{simulation, $\varepsilon = 0.914$};
   \addplot[dashed,line width=0.75pt,no markers,teal,domain=0.5:10.9,samples=200] {1.843*(x*x - 0.3674*x) / (x*x + 0.39070*x - 0.2516)};
   \addplot[dashed,line width=0.75pt,no markers,purple,domain=8.75:17,samples=200] {0.64*x^0.43};
%
  \node[] at (axis cs: 4.5,2.0) {\fcolorbox{teal}{white}{\scriptsize $ \dfrac{1.84 \,\left(x^2 - 0.37 \, x\right)}{x^2 + 0.39 \, x - 0.25}$}};
%
  \node[] at (axis cs: 14,2.3) {\fcolorbox{purple}{white}{\scriptsize $0.64 \, x^{0.43}$}};
  \end{axis}
\end{tikzpicture}
}
\caption{\zweizeilig Sherwood number as a function of the Reynolds number -- comparison of simulation and experiment. Symbols: simulation results; lines: fit formulae. For this plot, we adopt the definitions of Sherwood and Reynolds number provided in~\cite{Giani2005}, that is, for porosity $\varepsilon = 0.902$, $\text{Re}_\text{\Giani} = 0.204 \times \text{Re}$ and for $\varepsilon=0.914$, $\text{Re}_\text{\Giani} = 0.190 \times \text{Re}$. In this scaling, the function $\text{Sh}_\text{\Giani} = \text{Sh}_\text{\Giani} \left(\text{Re}_\text{\Giani}\right)$ is independent of the porosity, for details see~\cite{Giani2005}. The Schmidt number, $\text{Sc} = \nu / D_{CO} = 0.77$ is kept constant in this study.}

\label{fig:ShRe}
\end{figure}

Further, we wish to validate the model and algorithm by evaluating simulation data for the mass transport to the catalytic surface. Following Giani et al.~\cite{Giani2005}, we consider the Sherwood number as a function of the Reynolds number in the mass transfer limited regime. Numerically, this is achieved by using Eq.~\eqref{eq:reaction_prob} with
\begin{equation}
  \label{eq:3}
  \left< \frac{\text{d}n_{r}}{\text{d}t} \right> = \min\left(Z_\text{CO}, \, Z_{\text{H}_2\text{O}}\right)  \,.
\end{equation}
For a quantitative comparison of the simulation data with the experimental results from Giani et al., in Fig. \ref{fig:ShRe}, we normalize the simulation data to the data provided in~\cite{Giani2005}, as detailed in the figure caption. Hence, we define a Reynolds and Sherwood number based on the effective strut size~\cite{Giani2005},
\begin{equation}
  \label{eq:Res}
  \text{Re}_\text{\Giani} \equiv \dfrac{s U}{\nu}
\end{equation}
and
\begin{equation}
  \label{eq:Shs}
  \text{Sh}_\text{\Giani} = \dfrac{s}{D_{CO}} \, \dfrac{- \ln(1 - \eta)}{A_\text{s} \, h/U}  \, ,
\end{equation}
where $s$ is the effective strut size,
\begin{equation}
  \label{eq:strut}
  s = w \sqrt{\dfrac{4}{3 \pi} (1 - \varepsilon)}  \, .
\end{equation}
In the definition of the Sherwood number, $\eta$ denotes the conversion rate,
\begin{equation}
  \label{eq:conversion}
  \eta = 1 - \frac{\left.\frac{\text{d}n}{\text{d}t}\right|_\text{CO, out}} {\left.\frac{\text{d}n}{\text{d}t}\right|_\text{CO, in}}\,.
\end{equation}
within a system of length $h$. $D_{CO}$ and $A_\text{s}$ are the diffusion coefficient for carbon monoxide (the limiting educt species) and the specific surface of the foam, respectively.

The function $\text{Sh}_\text{\Giani} = \text{Sh}_\text{\Giani} \left(\text{Re}_\text{\Giani}\right)$ reveals two different regimes: For small Reynolds numbers, the data follows a rational function while for large Reynolds number they obey a power law~\cite{Giani2005}:
\begin{equation}
  \label{eq:4}
  \text{Sh}_\text{\Giani} = 
    \begin{cases}
      \dfrac{1.84 \,\left(\text{Re}_\text{\Giani}^2 - 0.37 \, \text{Re}_\text{\Giani} \right)}{\text{Re}_\text{\Giani} ^2 + 0.39 \, \text{Re}_\text{\Giani}  - 0.25} ~~ &\text{for} ~~\varepsilon\lesssim 10\\
      0.64 \, \text{Re}_\text{\Giani} ^{0.43} ~~~ &\text{for} ~~\varepsilon\gtrsim 10 
    \end{cases}
\end{equation}
This behavior can be explained as follows:
\subparagraph{Steep increase for small Re:}
The Sherwodd number, $ \text{Sh}_\text{\Giani} $, is directly proportional to the superficial velocity and increases monotonously with the conversion rate. 
%
Further, in addition to the convective transport, there is another process providing an inflow of reactants, namely the diffusive transport along the main flow direction. For small superficial velocity, the conversion is mainly fed by diffusion, and the conversion rate depends only weakly on the velocity. Therefore,  increasing $ \text{Re}_\text{\Giani} $, which is proportional to superficial velocity, leads to a steep increase of $\text{Sh}_\text{\Giani}$.
\subparagraph{Plateau for intermediate Re:}
For larger $ \text{Re}_\text{\Giani}$, however, the convective transport dominates, and $ \text{Sh}_\text{\Giani} $ approaches a plateau. 
To explain this plateau, we consider the probability for a particle to reach the catalytic surface, which is proportional to its residence time  and, therefore, inversely proportional to $U$. On the other hand, the influx of reactants is directly proportional to $U$, if we neglect diffusion along the main flow direction. Consequently, the product of both, that is, the mass transfer to the catalytic surface and, thus, the Sherwood number is nearly independent of the superficial velocity, leading to a plateau in Fig. \ref{fig:ShRe}.


\subparagraph{Power law for large Re:}
The explanation above does not hold true for $ \text{Re}_\text{\Giani} \gtrsim 10 $, and even before the mentioned above plateau is reached, the function turns into a power law with the exponent reported in~\cite{Giani2005}. While our simulation data are in very good agreement with the fit to the experimental data for the exponent, we underestimate the prefactor by Giani et al.~\cite{Giani2005} slightly.
We believe that the deviation should be attributed to the inverse sphere packing model of the foam which is more regular and less tortuous compared to real foam structures.

\subsection{Timescales of diffusion and reaction -- \Thiele{}}
\label{sec:timesc-diff-react}


An important dimensionless quantity for heterogeneous catalysis is the \Thiele{}~\cite{Thiele1939}, 
\begin{equation}
  \label{eq:Thiele}
  \Phi = \sqrt{\frac{\tau_d}{\tau_r }} \, ,
\end{equation}
where $ \tau_d $ and $ \tau_r $ are the diffusion and reaction timescales, respectively. For large  $\Phi$, the mass transfer to the catalytic surface is the limiting factor, while for small Thiele $\Phi$  the reaction rate limits the conversion~\cite{Deutschmann2009}. For the considered reaction, the bottleneck in the diffusion process is $\text{CO}$ as $\text{H}_2\text{O}$ is abundant in the considered system. Therefore, we define the diffusion time scale, $ \tau_d $, as the inverse of the diffusion rate per volume at which $\text{CO}$ is transported to the catalytic surface. Neglecting the duration of the reaction, that is, no $\text{CO}$ accumulates at the surface, the diffusion rate per volume is
\begin{equation}
  \label{eq:diffusion_time}
\frac{1}{\tau_d}= \frac{D_{CO} \, n_{\text{CO, in}}}{R^2} \, ,
\end{equation}
%
where $ n_\text{CO, in}$ is the number density of $\text{CO}$ at the inlet, and $R$ denotes the pore radius. Analogously, $\tau_r$ is the inverse of the reaction rate per volume. Assuming the diffusion to the reactive surface being much faster than the reaction, the reaction rate per volume is
\begin{equation}
  \label{eq:reaction_time}
\frac{1}{\tau_r} = \frac{\text{d} n_\text{r, in}}{\text{d}t} \, A_\text{s} \, L \, ,
\end{equation}
where $\text{d} n_\text{r, in}/\text{d}t$ is the average reaction rate in the washcoat, as given by Eq. \eqref{eq:reaction_rate} for inlet conditions. $A_\text{s}$ and $L$ are the specific surface and the washcoat thickness, respectively.

\begin{figure}[!htb]
\centering
\tikzsetnextfilename{Thiele}
 \resizebox{\figWidth}{!}{%
\begin{tikzpicture}
  \begin{loglogaxis}
  [
    axis line style={-},
    xlabel={$ \rho_{s} / \rho_{s,\text{ref}} $},
    ylabel={\Thiele{} $ \Phi $ [-]},
    xlabel near ticks,
    ylabel near ticks,
    xmin=1.0e-8, xmax=1.0,
    ymin=1.0e-2, ymax=1.0e2,
    tick style={color=black},
    tick label style={font=\footnotesize},
    xtick={0.00000001,0.0000001,0.000001,0.00001,0.0001,0.001,0.01,0.1,1},
    ytick={0.01,0.1,1,10,100},
    width=0.95\figWidth,
    height=0.7\figWidth,
  ]
   \addplot[only marks,line width=0.75pt,mark=o,mark size=2.0pt,red] table[x expr=(\thisrowno{0}), y expr=(\thisrowno{2}), col sep=space] {figs/Data/T453K_1atm_1090/Fitting_Thiele_modulus/Thiele_modulus.dat};
%
   \addplot[solid,line width=0.75pt,no markers,black,domain=1e-8:1e-3,samples=200] {10^(0.50*log10(x) + 2.14991)}
       coordinate [pos=0.3] (A) 
       coordinate [pos=0.5] (B);
%
   \addplot[solid,line width=0.75pt,no markers,black,domain=3.15e-5:1,samples=200] {10^(0.25*log10(x) + 1.2245)}
       coordinate [pos=0.45] (C) 
       coordinate [pos=0.65] (D);
%
   \draw (A) -| (B)  
      node [pos=0.25, anchor=north] {\scriptsize $1.0$} 
      node [pos=0.75,anchor=west]   {\scriptsize $0.5$}; 
   \draw (C) -| (D)  
      node [pos=0.25, anchor=north] {\scriptsize $1.0$} 
      node [pos=0.75,anchor=west]   {\scriptsize $0.25$}; 
  \end{loglogaxis}
\end{tikzpicture}
}
\caption{\zweizeilig \Thiele{} as a function of the normalized density of active sites. In agreement with literature~\cite{Roberts1966}, we find two characteristic regimes.}
\label{fig:Thiele}
\end{figure}
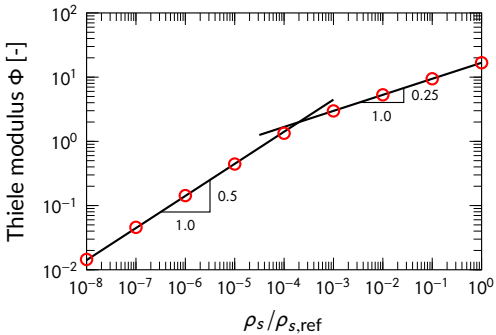

We study the \Thiele{} as a function of the normalized density of active reaction sites, $ \rho_{s} / \rho_{s,\text{ref}}$. The reference value, $\rho_{s,\text{ref}}$ used from normalization corresponds to the reaction parameters defined in the end of Sec. \ref{sec:wcParams}. We obtain
\begin{equation}
  \label{eq:5}
  \begin{aligned}
   \Phi &\propto  \left(\frac{\rho_{s}}{\rho_{s,\text{ref}}}\right)^{\frac{1}{2}} & \text{for}~~~ \frac{\rho_{s}}{\rho_{s,\text{ref}}}\lesssim 3\times 10^{-4}\\
   \Phi &\propto  \left(\frac{\rho_{s}}{\rho_{s,\text{ref}}}\right)^{\frac{1}{4}} & \text{for}~~~ \frac{\rho_{s}}{\rho_{s,\text{ref}}}\gtrsim 3\times 10^{-4}
   \end{aligned}
\end{equation}
as shown in Fig. \ref{fig:Thiele}.

For low densities of active sites,  the \Thiele{} grows with the square root of the normalized density of active sites and, consequently, the average reaction rate in the washcoat increases linearly  with the relative active site density. This indicates that the reactions take place homogeneously within the washcoat, that is, $\zeta = 1$. 
For larger active site density, the exponent changes to  1/4, indicating that the effectiveness factor decreases with the exponent 0.5. The results of the simulation coincides with the findings by Roberts and Satterfield~\cite{Roberts1966}.

\section{Simulation-based predictions for reactive flow in porous media}
\label{sec:simul-based-pred}

Based on the excellent quantitative agreement of our simulation results with experimental data from different sources of the literature, shown in Secs. \ref{sec:pressure-drop-hagen}-\ref{sec:timesc-diff-react} we feel confident that our numerical method and the corresponding models deliver a correct and reliable description of heterocatalytic reactive flows. Particularly good agreement we found for flows of low Reynolds number. Therefore, in the subsequent Secs. \ref{sec:active-site-density}-\ref{sec:porosity}, we will exploit our method's predictive power to study the dependence of the system on the density of active sites in the catalyst and on the foam porosity, in the regime of low Reynolds number. Both parameters are difficult to vary and, thus, not easily accessible in experimental investigations. Having in mind applications like microfluidics, where due to small system sizes, also the \Thiele{} is typically small, we extend the study of the mass transfer limited regime also to the intermediate range towards the reaction rate limited regime.

\subsection{Simulation setup}
\label{sec:simul-setup}

\subsubsection{Bulk transport coefficients}
\label{sec:fluidProps}

Typical mole fractions of $\text{CO}$, $\text{CO}_2$, $\text{H}_2$, $\text{N}_2$, and $\text{H}_2\text{O}$ for the low temperature water-gas shift reaction at the inlet are~\cite{Ratnasamy2009}, $3 \, \%$, $13 \, \%$, $30 \, \%$, $28 \, \%$, and $26 \, \%$, respectively. Characteristic temperature and pressure are $ T = 453 \, \text{K} $ and $ P = 1 \, \text{atm} $, respectively. We obtain the binary diffusion coefficient for each pair of species using Hirschfelder's method~\cite{Hirschfelder1954} and subsequently the bulk diffusion coefficients for each species in the mixture are computed~\cite{Fairbanks1950}. For the gas mixture at the inlet, we obtain
\begin{equation}
  \label{eq:6}
  \begin{split}
D_\text{CO} & = 5.842 \times 10^{-5} \, \text{m}^2 / \text{s}\\
D_{\text{H}_2\text{O}} & = 8.795 \times 10^{-5} \, \text{m}^2 / \text{s}\\
D_{\text{CO}_2} &= 5.225 \times 10^{-5} \, \text{m}^2 / \text{s}\\
D_{\text{H}_2} & = 1.682 \times 10^{-4} \, \text{m}^2 / \text{s}
\end{split}
\end{equation}
As the reaction proceeds, the diffusion coefficients change with the composition of the gas mixture. Assuming that $\text{CO}$ is completely consumed due to the reaction, the diffusion coefficient for $\text{CO}$, $\text{H}_2\text{O}$, $\text{CO}_2$ and $\text{H}_2$ would change by $ -0.9 \, \% $, $ +1.2 \, \% $, $ -4.2 \, \% $, and $ 1.9 \, \% $, respectively. We neglect this effect here.

For the computation of the viscosity of the gas mixture, according to~\cite{Buddenberg1949}, we need the mass density and the viscosity of each involved component. The mass densities are computed from the ideal gas law. The viscosity of overheated steam is computed according to IAPWS-IF97~\cite{Wagner2008}, while the dynamic viscosities of the remaining components are computed as described in~\cite{Sutherland1893}. From the densities and the dynamic viscosities,
\begin{equation}
  \label{eq:7}
 \begin{aligned}
\rho_\text{CO} & = 0.7535 \, \text{kg} / \text{m}^3    & \mu_\text{CO} = 2.412 \times 10^{-5} \, \text{Pa} \, \text{s} \\
\rho_{\text{H}_2\text{O}} &= 0.4846 \, \text{kg} / \text{m}^3 & \mu_{\text{H}_2\text{O}} = 1.537 \times 10^{-5} \, \text{Pa} \, \text{s}\\
\rho_{\text{CO}_2} &= 1.184 \, \text{kg} / \text{m}^3   & \mu_{\text{CO}_2} = 2.188 \times 10^{-5} \, \text{Pa} \, \text{s} \\
\rho_{\text{H}_2} &= 0.05423 \, \text{kg} / \text{m}^3 & \mu_{\text{H}_2} = 1.169 \times 10^{-5} \, \text{Pa} \, \text{s} \\
\rho_{\text{N}_2} &= 0.7536 \, \text{kg} / \text{m}^3   & \mu_{\text{N}_2} = 2.406 \times 10^{-5} \, \text{Pa} \, \text{s}
\end{aligned}
\end{equation}
and the mole fractions given above, we compute the average density, $ \rho = 0.5298 \, \text{kg} / \text{m}^3 $, and the average dynamic viscosity, $ \mu = 2.368 \times 10^{-5} \, \text{Pa} \, \text{s} $.

\subsubsection{Simulation parameters}
\label{sec:simParams}

In particle-based, mesoscopic simulation methods, the fluid properties are represented by quasi-particles. Further, the transport coefficients of the simulated fluid depend on the spatial and temporal discretization as well as some additional parameters~\cite{Ihle2003a, Ihle2003b, Kikuchi2003, Muehlbauer2017}. In practical cases, the parameters of the fluid model, such as, rotation angle (in SRD and iSRD), collision volume size, or time step, must be chosen in such a way that the dimensionless characteristics of the real system, such as Reynolds number, P\'eclet number, and Schmidt number, match the corresponding characteristics in the real system. 
Here we describe one way to determine suitable simulation parameters.

As for most particle-based methods, for iSRD the transport coefficients depend on the spatial and temporal discretization. Moreover, for iSRD, the diffusion coefficient of a particle is directly related to its mass: For fixed density and rotation angle, $ D \propto \Delta t \, k_B T / m $ ~\cite{Muehlbauer2017}. In order to avoid an additional thermostat to be necessary for the inlet-outlet boundary condition, we choose $ m_{\text{CO, sim}} = m_{\text{CO}_{2},\text{sim}} $ and $ m_{\text{H}_{2}\text{O},\text{sim}} = m_{\text{H}_{2},\text{sim}} $, which implies $ D_\text{CO, sim} = D_{\text{CO}_{2},\text{sim}} $ and $ D_{\text{H}_{2}\text{O},\text{sim}} = D_{\text{H}_{2},\text{sim}} $. This simplification has only minor influence on the simulation results, since the transport away from the reactive surface is secondary.

The ratio between diffusion coefficient and viscosity can be tuned via the rotation angle and the ratio between time step and collision diameter~\cite{Muehlbauer2017}. For the chosen discretization, $ D_{\text{CO}, \text{sim}} $ and $\nu_{\text{sim}}$ are $ \sigma = 27.0 $ times larger than the values given in Tab. \ref{tab:setup}.
\begin{table*}[ht]
\caption{\zweizeilig Summary of the dimensional and dimensionless parameters relevant for the following studies.}
\begin{center}
  \begin{threeparttable}
    \begin{tabular}{l lr}
      \headrow
  \textbf{dimensional} \hspace*{10mm}   &   & \textbf{value} \\
  \hline
  kinematic viscosity                   & $ \nu $                                        & $ 4.469 \times 10^{-5} \, \text{m}^2 / \text{s} $ \\
  diffusion coefficient                 & $ D_{CO} $                                     & $ 5.842 \times 10^{-5} \, \text{m}^2 / \text{s} $ \\
  prism width                           & $ w $                                          & $ 0.635 \times 10^{-3} $ m \\
  prism height                          & $ h = \sqrt{8/3} \, w $                        & $ 1.037 \times 10^{-3} $ m \\
  superficial velocity                  & $ U $                                          & $ 1.0 $ m$/$s \\
      \hline
      \headrow
  \textbf{dimensionless}                &   & \textbf{value} \\
  Reynolds number                       & $ \text{Re} = w \, U / \nu $                 & $ 14.2 $ \\
  P\'{e}clet number                     & $ \text{Pe} = w \, U / D_{CO} $              & $ 10.9 $ \\
  Schmidt number                        & $ \text{Sc} = \nu / D_{CO} $                 & $ 0.77 $ \\
\end{tabular}
\end{threeparttable}
\end{center}
\label{tab:setup}
\end{table*}
To compensate for that, the time scale of convective transport as well as the reaction time scale in the unit cell can be chosen appropriately. To this end, both the superficial velocity, $ U $, and the effective reaction rate, $ \langle \dot{n}_{r} \rangle $, must be increased by the same factor $ \sigma $. This approach ensures that, while the \emph{dimensional} parameters chosen in the simulations may deviate from what is shown in Tab. \ref{tab:setup}, all relevant \emph{dimensionless} parameters -- Reynolds number, Schmidt number and Thiele{} -- match the actual application. Note that for a given Reynolds and Schmidt number also the P\'{e}clet number is known. The temperature is assumed to be constant within the simulation domain. Therefore, all dimensionless numbers that refer to heat transfer are irrelevant for the considered setup.


\subsubsection{System geometry and boundary conditions}
\label{sec:syst-geom-bound}

\begin{figure}
	\centering
	\includegraphics[width=\figWidth]{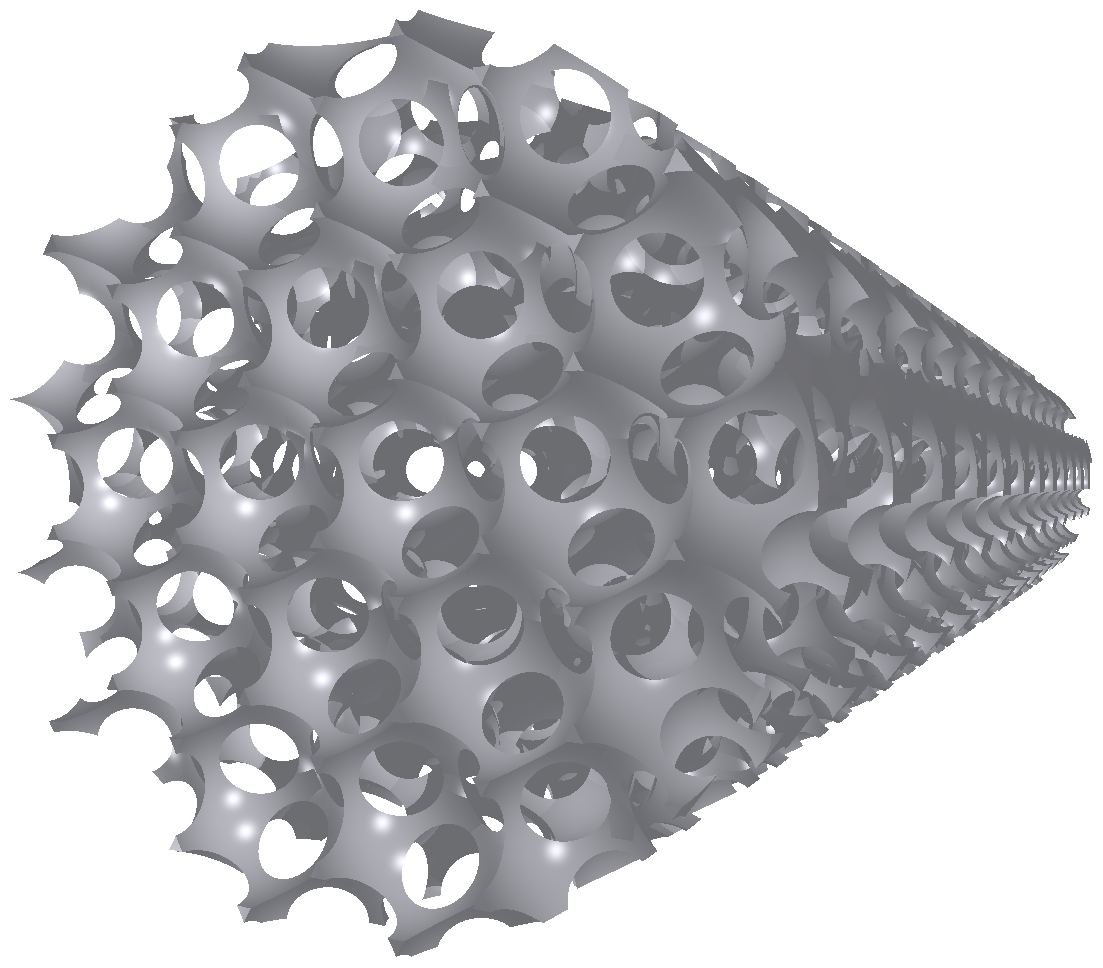}

	\caption{\zweizeilig Cylindrical cutout of a foam structure modeled as an inverse sphere packing.}
	\label{fig:FoamSev}
\end{figure}

We are interested in foam structures, as shown in Figure~\ref{fig:FoamSev}. In the simulation, we consider representative foam cells. 
Figure~\ref{fig:Foam} shows the stream lines and the field of mole fraction,
\begin{equation}
  \label{eq:concentration}
  {x}_i = \dfrac{{n}_{i}}{\sum_j {n}_{j}} \, ,
\end{equation}
for a typical simulation setup. At the inlet, the concentration is chosen homogeneous. 

In order to vary the porosity, we adjust the diameter of the spheres used for the inverse sphere packing described in Sec. \ref{sec:foam} while keeping the width of the foam cell constant. At the same time, the superficial velocity is kept constant.

For practical applications one aims for foam substrates for heterogeneous catalysis which combine a large conversion rate, $\eta$, with a small pressure drop, $ \Delta P $. The conversion rate, $\eta$, describes the amount of educt which is converted along the reactor. Naturally, this quantity depends on the system's length. Assuming an exponential decay of the reactant concentration between inlet (position: $z = 0$) and outlet (position: $z = h$),
\begin{equation}
  \label{eq:expDecay}
  \left.\frac{\text{d}n}{\text{d}t}\right|_{\text{CO, } z} = \left.\frac{\text{d}n}{\text{d}t}\right|_\text{CO, in} \, \exp \left(-\langle \eta \rangle \, \dfrac{z}{h} \right) \, ,
\end{equation}
an effective conversion rate,
\begin{equation}
  \label{eq:effConversion}
  \langle \eta \rangle = -\ln(1 - \eta) \, ,
\end{equation}
can be defined, which is independent of the channel length.

\subsection{Active site density}
\label{sec:active-site-density}

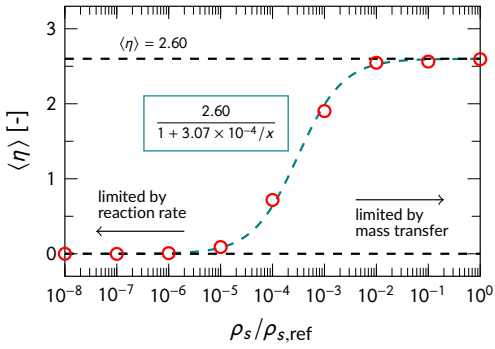
\begin{figure}[!htb]
\centering
\tikzsetnextfilename{WashcoatActivityEff}
 \resizebox{\figWidth}{!}{%
\begin{tikzpicture}
  \begin{semilogxaxis}
  [
    axis line style={-},
    xlabel={$ \rho_{s} / \rho_{s,\text{ref}} $},
    ylabel={$ \langle \eta \rangle $ [-]},
    xlabel near ticks,
    ylabel near ticks,
    xmin=1.0e-8, xmax=1.0,
    ymin=-0.3, ymax=3.3,
    tick style={color=black},
    tick label style={font=\footnotesize},
    xtick={0.00000001,0.0000001,0.000001,0.00001,0.0001,0.001,0.01,0.1,1},
    minor y tick num={1},
    width=0.95\figWidth,
    height=0.7\figWidth,
  ]
   \addplot[dashed,line width=0.75pt,no markers,teal,domain=1e-8:1,samples=200] { 2.60/(1 + 10^(- 3.51244) * x^(-1.0))};
   \addplot[only marks,line width=0.75pt,mark=o,mark size=2.0pt,red] table[x expr=(\thisrowno{19}), y expr=(-ln(\thisrowno{18} / (0.03 * \thisrowno{7}))), col sep=space] {figs/Data/T453K_1atm_1090/results_vary_rrc_FINAL.dat};
%
   \addplot[dashed,line width=0.75pt,black,domain=1e-8:1] {0.0};
%
   \addplot[dashed,line width=0.75pt,black,domain=1e-8:1] {2.60};
%
  \draw[->] (axis cs: 4e-3,0.72)--(axis cs: 2e-1,0.72);
  \draw[->] (axis cs: 2e-6,0.3)--(axis cs: 4e-8,0.3);
%
  \node[] at (axis cs: 5e-7,2.8) {\scriptsize $ \langle \eta \rangle = 2.60$};
  \node[anchor=south east] at (axis cs: 3e-4,1.25) {\fcolorbox{teal}{white}{\scriptsize $\dfrac{2.60}{1 +  3.07 \times 10^{-4} / x}$}};
  \node[anchor=south west] at (axis cs: 2.7e-3,0.30) {\scriptsize limited by};
  \node[anchor=south west] at (axis cs: 2.7e-3,0.12) {\scriptsize mass transfer};
  \node[anchor=south west] at (axis cs: 3e-8,0.54) {\scriptsize limited by};
  \node[anchor=south west] at (axis cs: 3e-8,0.36) {\scriptsize reaction rate};
  \end{semilogxaxis}
%
%
\end{tikzpicture}
}
\caption{\zweizeilig For the fixed porosity value $\varepsilon = 0.902$, we vary the relative active site density in the washcoat. The dashed line has been fitted to the data.}
\label{fig:washcoat_activity_eff}
\end{figure}

First we study the effective conversion rate, $\langle \eta \rangle$, as a function of the active site density in the washcoat. Changing this parameter allows to control the mode of operation of the system from the reaction rate limited regime to the  mass transfer limited regime. Figure~\ref{fig:washcoat_activity_eff} shows the conversion rate over the density of active sites, normalized by the density of active sites in the reference system according to the reaction parameters proposed by Aya\-stuy et al.~\cite{Ayastuy2005} and the geometric washcoat properties described by Novak et al.~\cite{Novak2012} (see Sec.~\ref{sec:wcParams}).
Clearly, the reference system, corresponding to $\rho_{s} / \rho_{s,\text{ref}}=1$, operates in the mass transfer limited regime. Starting from the reference value, we gradually reduce the active site density to drive the system towards the reaction rate limited regime.  Table~\ref{tab:Thiele} and Figure~\ref{fig:Thiele} show the \Thiele{} as a function of the relative site density.
\begin{table}[ht]
\caption{\zweizeilig \Thiele{}, $ \Phi $, as function of the active site density over reference site density, $ \rho_{s} / \rho_{s,\text{ref}} $.}
\begin{center}
\begin{threeparttable}
  \begin{tabular}{l l}
    \headrow
    $ \rho_{s} / \rho_{s,\text{ref}} $ & $ \Phi $\\
    $ 10^{-8} $  & $0.0145$\\
    $ 10^{-7} $  & $0.0458$\\
    $ 10^{-6} $  & $0.143$ \\
    $ 10^{-5} $  & $0.443$\\
    $ 10^{-4} $  & $1.34$\\
    $ 10^{-3} $  & $2.98$\\
    $ 10^{-2} $  & $5.31$\\
    $ 10^{-1} $  & $9.44$ \\
    $ 10^{0} $  & $16.8$
\end{tabular}
\end{threeparttable}
\end{center}
\label{tab:Thiele}
\end{table}

Even in the considered case, where the Reynolds and P\'{e}clet numbers are at the lower end of the industrially relevant range, the active site density, and such, the total amount of catalyst, may be reduced by the factor $100$ compared to the reference configuration~\cite{Ayastuy2005} without notably reducing the conversion rate. Further, half of the maximum effective conversion is reached at a relative site density of $ 3.07 \times 10^{-4} $. This corresponds to $ \Phi \approx 2 $, which is also where the two regimes meet in Fig. \ref{fig:Thiele}.

\subsection{Porosity study}
\label{sec:porosity}

In this section, we will perform a porosity study for two different values of the active site density:
\begin{itemize}
 \setlength\itemsep{0mm}
 \item[\textasteriskcentered] \makebox[4.4cm]{reference case (based on~\cite{Ayastuy2005,Novak2012}) \hfill} $\rightarrow$ $\Phi = 16.8$,
 \item[\textasteriskcentered] \makebox[4.4cm]{active sites reduced to $0.01 \, \%$ \hfill}                   $\rightarrow$ $\Phi = 1.34$.
\end{itemize}
The reference case represents the mass transfer limited regime, while the other configuration yields an intermediate regime, as can be seen in Figure~\ref{fig:washcoat_activity_eff}.

\begin{figure}[!htb]
\centering
 \resizebox{\figWidth}{!}{%
\tikzsetnextfilename{PressureDrop}
\begin{tikzpicture}
  \begin{axis}
  [
    axis line style={-},
    xlabel={porosity [-]},
    ylabel={$ \Delta P / (\rho \, U^2) $ [-]},
    xlabel near ticks,
    ylabel near ticks,
    xmin=0.86, xmax=0.94,
    ymin=8, ymax=20,
    tick style={color=black},
    tick label style={font=\footnotesize},
    x tick label style={
      /pgf/number format/.cd,
      fixed,
      fixed zerofill,
      precision=2,
      /tikz/.cd },
    xtick={0.86,0.87,...,0.94},
    ytick={0,2,...,20},
    minor x tick num={1},
    minor y tick num={1},
    width=0.95\figWidth,
    height=0.60\figWidth,
  ]
  \addplot[line width=0.75pt,mark=o,mark size=2.0pt,red] table[x expr=(\thisrowno{2}), y expr=(\thisrowno{13} / (\thisrowno{8} * \thisrowno{11} * \thisrowno{11})), col sep=space] {figs/Data/T453K_1atm_1e-0rrc/results_vary_k_FINAL.dat};
%
  \addplot[dashed, samples=10, smooth,domain=0:1,black, name path=one] coordinates {(0.872941,4)(0.872941,20)};
  \addplot[dashed, samples=10, smooth,domain=0:1,black, name path=two] coordinates {(0.901724,4)(0.901724,20)};
  \addplot[dashed, samples=10, smooth,domain=0:1,black, name path=three] coordinates {(0.926152,4)(0.926152,20)};
  \end{axis}
  \node at (0.9,3.1){
    \includegraphics[width=0.15\figWidth]{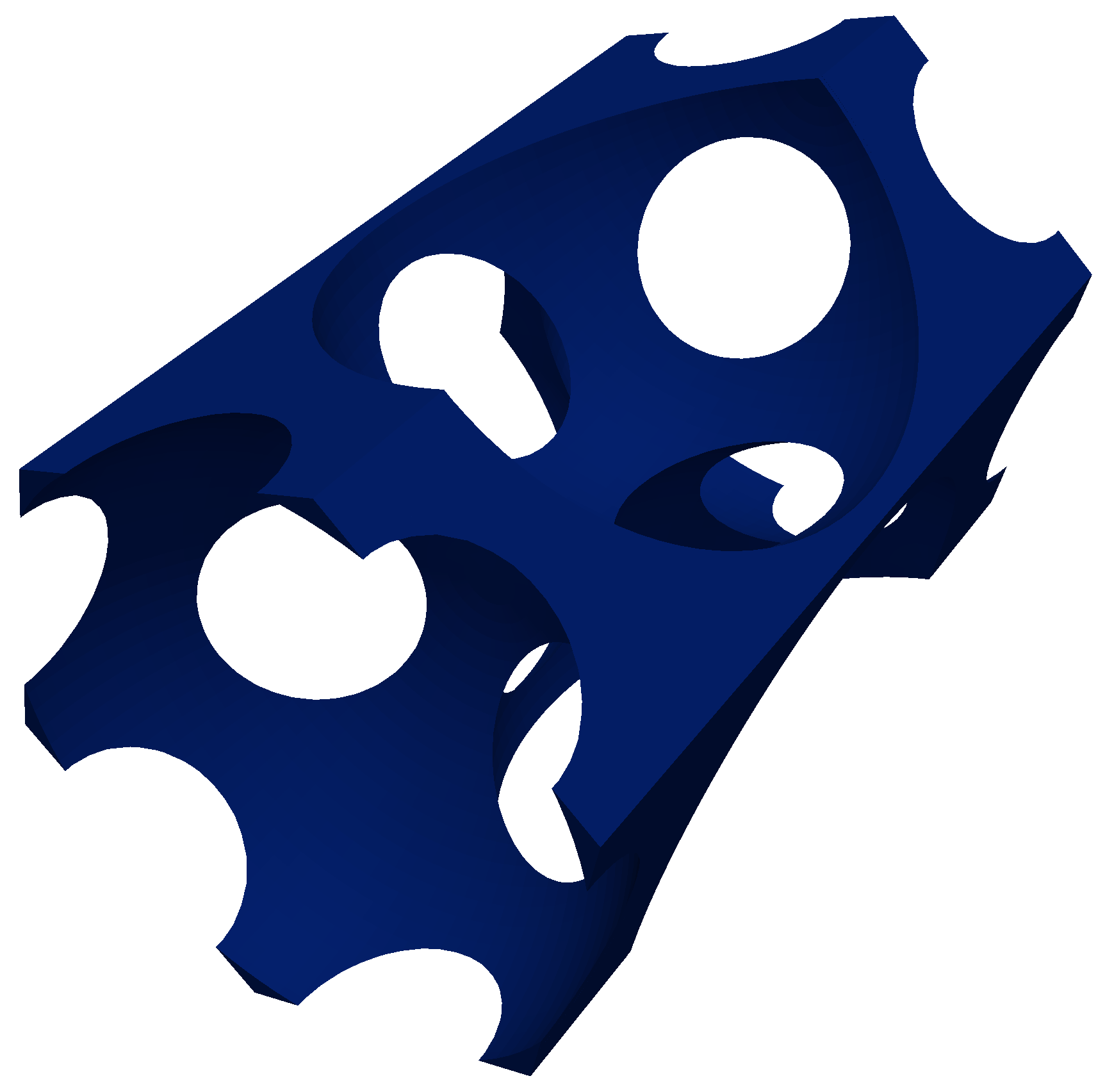}
  };
  \node at (2.65,3.1){
    \includegraphics[width=0.15\figWidth]{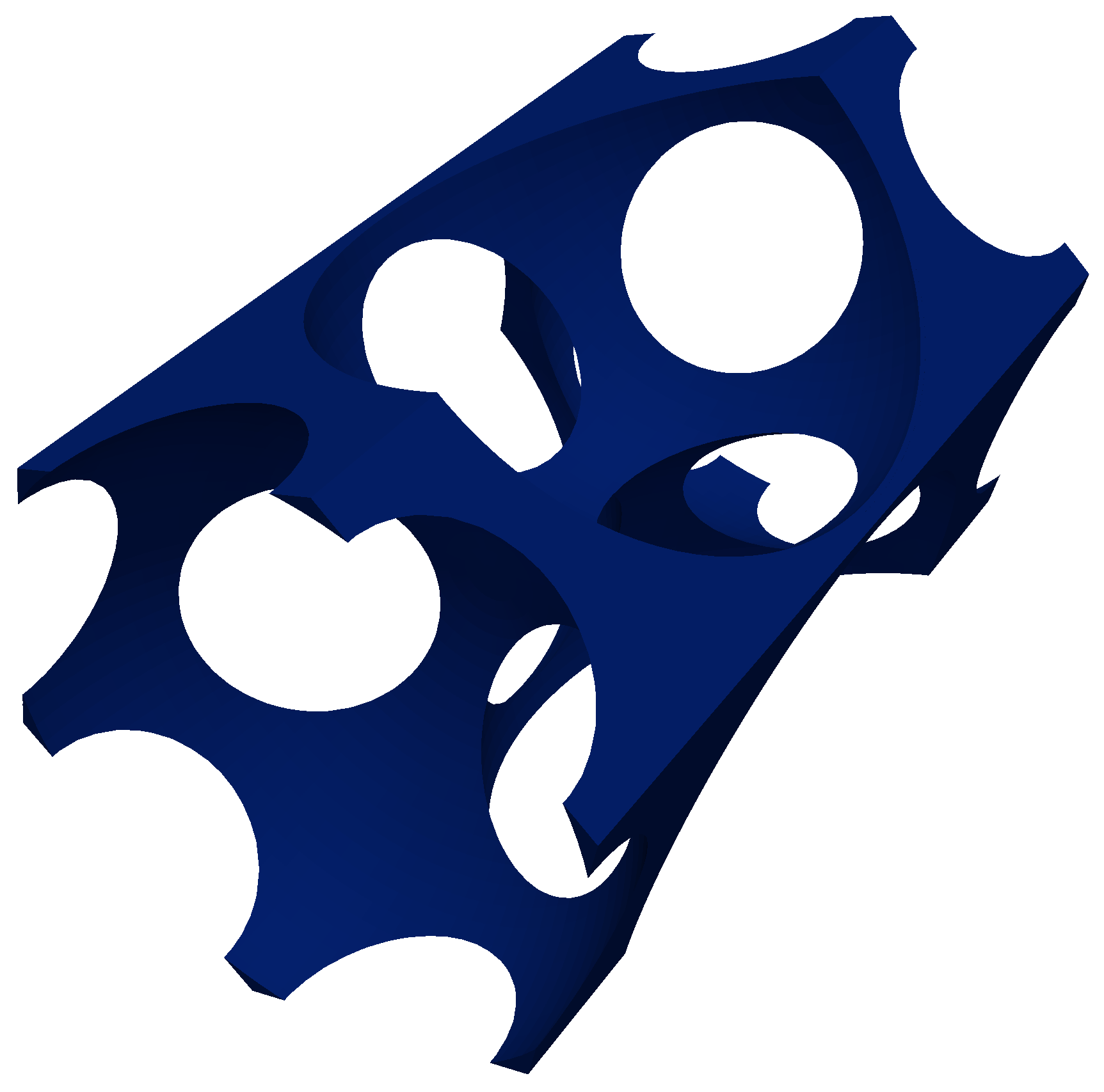}
  };
  \node at (4.15,3.1){
    \includegraphics[width=0.15\figWidth]{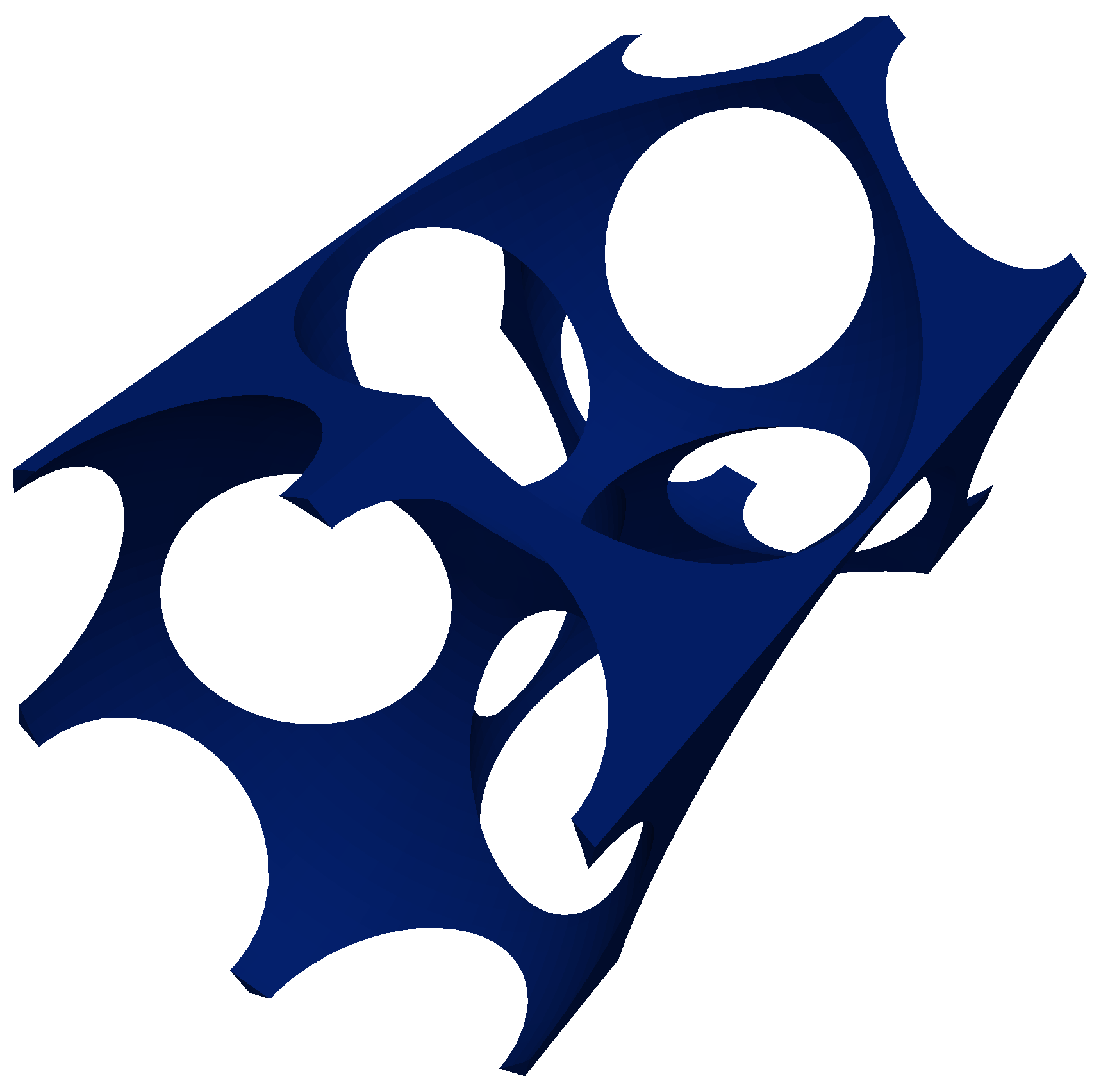}
  };
\end{tikzpicture}
}
\caption{\zweizeilig For fixed mass flow rate, we plot the dimensionless pressure drop as a function of porosity. The flow resistance decreases with increasing porosity.}
\label{fig:pressure_drop}
\end{figure}
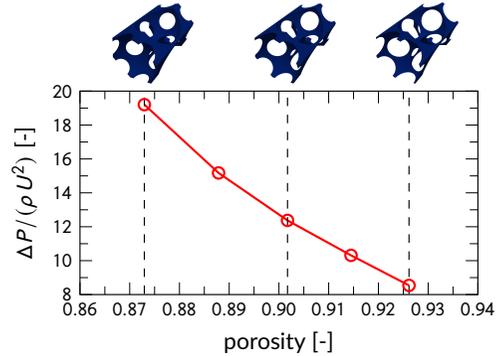
\begin{figure}[!htb]
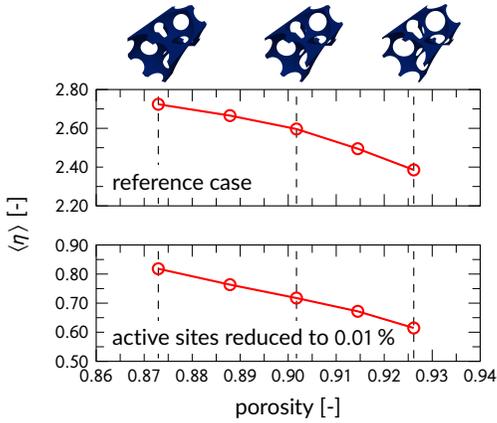

\centering
\tikzsetnextfilename{ConversionRate}
 \resizebox{\figWidth}{!}{%
\begin{tikzpicture}
  \begin{groupplot}
  [
    group style={group size=1 by 2,vertical sep= 0.5cm,horizontal sep= 2.0cm},
    xlabel near ticks,
    ylabel near ticks,
    xmin=0.86, xmax=0.94,
    tick style={color=black},
    tick label style={font=\footnotesize},
    minor x tick num={1},
    minor y tick num={1},
    width=0.95\figWidth,
    height=0.45\figWidth,
    ]
    \nextgroupplot
    [
    axis line style={-},
    xmin=0.86, xmax=0.94,
    ymin=2.2, ymax=2.8,
    y tick label style={
      /pgf/number format/.cd,
      fixed,
      fixed zerofill,
      precision=2,
      /tikz/.cd },
    xtick={0.86,0.87,...,0.94},
    xticklabels={},
    ]
      \addplot[dashed, samples=10, smooth,domain=0:1,black, name path=one] coordinates {(0.872941,0)(0.872941,3.0)};
      \addplot[dashed, samples=10, smooth,domain=0:1,black, name path=two] coordinates {(0.901724,0)(0.901724,3.0)};
      \addplot[dashed, samples=10, smooth,domain=0:1,black, name path=three] coordinates {(0.926152,0)(0.926152,3.0)};
      \node[anchor=south west] (label1) at (axis cs: 0.86,2.20) {\colorbox{white}{reference case}};
      \addplot[line width=0.75pt,mark=o,mark size=2.0pt,red] table[x expr=(\thisrowno{2}), y expr=(-\thisrowno{17}), col sep=space] {figs/Data/T453K_1atm_1e-0rrc/results_vary_k_FINAL.dat};
%
    \nextgroupplot
    [
    axis line style={-},
    xlabel={porosity [-]},
    xmin=0.86, xmax=0.94,
    ymin=0.50, ymax=0.9,
    x tick label style={
      /pgf/number format/.cd,
      fixed,
      fixed zerofill,
      precision=2,
      /tikz/.cd },
    y tick label style={
      /pgf/number format/.cd,
      fixed,
      fixed zerofill,
      precision=2,
      /tikz/.cd },
    xtick={0.86,0.87,...,0.94},
    ]
      \addplot[dashed, samples=10, smooth,domain=0:1,black, name path=one] coordinates {(0.872941,0)(0.872941,1.0)};
      \addplot[dashed, samples=10, smooth,domain=0:1,black, name path=two] coordinates {(0.901724,0)(0.901724,1.0)};
      \addplot[dashed, samples=10, smooth,domain=0:1,black, name path=three] coordinates {(0.926152,0)(0.926152,1.0)};
      \node[anchor=south west] (label2) at (axis cs: 0.86,0.50) {\colorbox{white}{active sites reduced to $0.01 \, \%$}};
      \addplot[line width=0.75pt,mark=o,mark size=2.0pt,red] table[x expr=(\thisrowno{2}), y expr=(-\thisrowno{17}), col sep=space] {figs/Data/T453K_1atm_1e-4rrc/results_vary_k_FINAL.dat};
%
  \end{groupplot}
  \node at (0.9,2.075){
    \includegraphics[width=0.15\figWidth]{figs/BasisCell_Examples/foam-1070.png}
  };
  \node at (2.65,2.075){
    \includegraphics[width=0.15\figWidth]{figs/BasisCell_Examples/foam-1090.png}
  };
  \node at (4.15,2.075){
    \includegraphics[width=0.15\figWidth]{figs/BasisCell_Examples/foam-1110.png}
  };
  \node (y_label) at ($(group c1r1.north west)!0.5!(group c1r2.south west)$) [xshift=-1.0cm] {\rotatebox{90}{$\langle \eta \rangle$ [-]}};
%
\end{tikzpicture}
}
\caption{\zweizeilig For fixed mass flow rate, we plot the effective conversion rate as a function of porosity. The conversion decreases with increasing porosity.}
\label{fig:conversion_rate}
\end{figure}
As Figures~\ref{fig:pressure_drop} and ~\ref{fig:conversion_rate} show, both dimensionless pressure drop and conversion rate decline, as the porosity is increased. This is expected, since with increasing porosity, the strut size as well as the surface area decreases. For the performed porosity variation, in the mass transfer limited regime, the effective conversion rate varies in the range $\left[ -8 \, \% \, , +5 \, \% \right]$ around the value for the median porosity, $\varepsilon = 0.902$. In the intermediate regime, the effective conversion rate varies in the range $\left[ -15 \, \% \, , +14 \, \% \right]$. The variation is more distinctive in the intermediate regime, since there the reactants need more wall encounters on average, before the reaction finally takes place.

\begin{figure}[!htb]
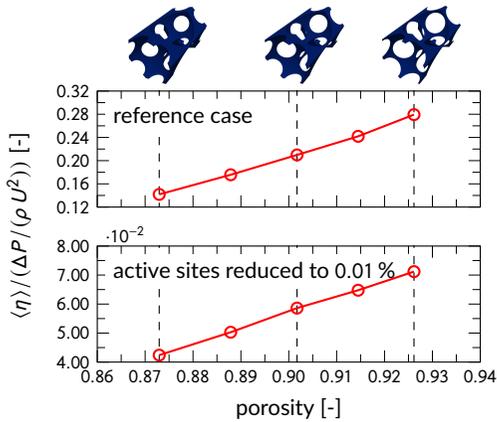

\centering
\tikzsetnextfilename{ObjectiveFunction}
 \resizebox{\figWidth}{!}{%
\begin{tikzpicture}
  \begin{groupplot}
  [
    group style={group size=1 by 2,vertical sep= 0.5cm,horizontal sep= 2.0cm},
    xlabel near ticks,
    ylabel near ticks,
    xmin=0.86, xmax=0.94,
    tick style={color=black},
    tick label style={font=\footnotesize},
    minor x tick num={1},
    minor y tick num={1},
    width=0.95\figWidth,
    height=0.45\figWidth,
    ]
    \nextgroupplot
    [
    axis line style={-},
    xmin=0.86, xmax=0.94,
    ymin=0.12, ymax=0.32,
    y tick label style={
      /pgf/number format/.cd,
      fixed,
      fixed zerofill,
      precision=2,
      /tikz/.cd },
    xtick={0.86,0.87,...,0.94},
    xticklabels={},
    ytick={0.0,0.04,...,0.40},
    ]
      \addplot[dashed, samples=10, smooth,domain=0:1,black, name path=one] coordinates {(0.872941,0)(0.872941,0.40)};
      \addplot[dashed, samples=10, smooth,domain=0:1,black, name path=two] coordinates {(0.901724,0)(0.901724,0.40)};
      \addplot[dashed, samples=10, smooth,domain=0:1,black, name path=three] coordinates {(0.926152,0)(0.926152,0.40)};
      \node[anchor=north west] (label1) at (axis cs: 0.86,0.32) {\colorbox{white}{reference case}};
      \addplot[line width=0.75pt,mark=o,mark size=2.0pt,red] table[x expr=(\thisrowno{2}), y expr=(-\thisrowno{17} / (\thisrowno{13} / (\thisrowno{8} * \thisrowno{11} * \thisrowno{11}))), col sep=space] {figs/Data/T453K_1atm_1e-0rrc/results_vary_k_FINAL.dat};
%
    \nextgroupplot
    [
    axis line style={-},
    xlabel={porosity [-]},
    xmin=0.86, xmax=0.94,
    ymin=0.04, ymax=0.08,
    x tick label style={
      /pgf/number format/.cd,
      fixed,
      fixed zerofill,
      precision=2,
      /tikz/.cd },
    y tick label style={
      /pgf/number format/.cd,
      fixed,
      fixed zerofill,
      precision=2,
      /tikz/.cd },
    xtick={0.86,0.87,...,0.94},
    ytick={0.0,0.01,...,0.10},
    ]
      \addplot[dashed, samples=10, smooth,domain=0:1,black, name path=one] coordinates {(0.872941,0)(0.872941,0.10)};
      \addplot[dashed, samples=10, smooth,domain=0:1,black, name path=two] coordinates {(0.901724,0)(0.901724,0.10)};
      \addplot[dashed, samples=10, smooth,domain=0:1,black, name path=three] coordinates {(0.926152,0)(0.926152,0.10)};
      \node[anchor=north west] (label2) at (axis cs: 0.86,0.080) {\colorbox{white}{active sites reduced to $0.01 \, \%$}};
      \addplot[line width=0.75pt,mark=o,mark size=2.0pt,red] table[x expr=(\thisrowno{2}), y expr=(-\thisrowno{17} / (\thisrowno{13} / (\thisrowno{8} * \thisrowno{11} * \thisrowno{11}))), col sep=space] {figs/Data/T453K_1atm_1e-4rrc/results_vary_k_FINAL.dat};
%
  \end{groupplot}
  \node at (0.9,2.075){
    \includegraphics[width=0.15\figWidth]{figs/BasisCell_Examples/foam-1070.png}
  };
  \node at (2.65,2.075){
    \includegraphics[width=0.15\figWidth]{figs/BasisCell_Examples/foam-1090.png}
  };
  \node at (4.15,2.075){
    \includegraphics[width=0.15\figWidth]{figs/BasisCell_Examples/foam-1110.png}
  };
  \node (y_label) at ($(group c1r1.north west)!0.5!(group c1r2.south west)$) [xshift=-1.0cm] {\rotatebox{90}{$ \langle \eta \rangle / (\Delta P / (\rho \, U^2)) $ [-]}};
%
\end{tikzpicture}
}
\caption{\zweizeilig To evaluate the suitability for the use as substrate in heterogeneous catalysis, we consider the performance index suggested by Giani et al. in~\cite{Giani2005}.}
\label{fig:objectiveFunction}
\end{figure}
An optimum substrate for heterogeneous catalysis should provide maximum conversion rate, while causing minimum pressure drop. However, these two requirements cannot be fulfilled at the same time. To find a compromise, the performance index as in~\cite{Giani2005} is evaluated for varying porosities in Figure~\ref{fig:objectiveFunction}. In both considered cases, the performance index monotonously increases with porosity. Note that -- due to the stronger relative decline in $\langle \eta \rangle$ depicted in the lower half of Figure~\ref{fig:conversion_rate} -- the increase of the performance index is less distinct in the intermediate regime.

\begin{figure}[!htb]
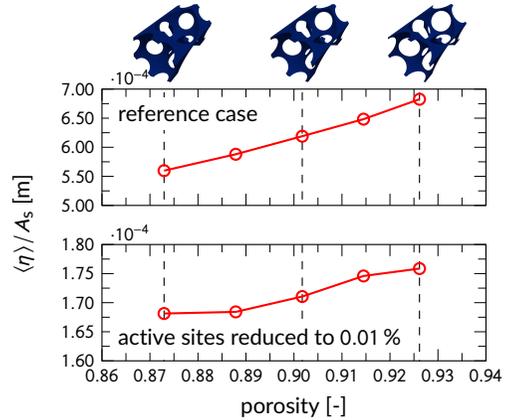

\centering
\tikzsetnextfilename{SurfaceUtilization}
 \resizebox{\figWidth}{!}{%
\begin{tikzpicture}
  \node at (0.9,2.075){
    \includegraphics[width=0.15\figWidth]{figs/BasisCell_Examples/foam-1070.png}
  };
  \node at (2.65,2.075){
    \includegraphics[width=0.15\figWidth]{figs/BasisCell_Examples/foam-1090.png}
  };
  \node at (4.15,2.075){
    \includegraphics[width=0.15\figWidth]{figs/BasisCell_Examples/foam-1110.png}
  };
  \begin{groupplot}
  [
    group style={group size=1 by 2,vertical sep= 0.5cm,horizontal sep= 2.0cm},
    xlabel near ticks,
    ylabel near ticks,
    xmin=0.86, xmax=0.94,
    tick style={color=black},
    tick label style={font=\footnotesize},
    minor x tick num={1},
    minor y tick num={1},
    width=0.95\figWidth,
    height=0.45\figWidth,
    ]
    \nextgroupplot
    [
    axis line style={-},
    xmin=0.86, xmax=0.94,
    ymin=5.0e-4, ymax=7.0e-4,
    y tick label style={
      /pgf/number format/.cd,
      fixed,
      fixed zerofill,
      precision=2,
      /tikz/.cd },
    xtick={0.86,0.87,...,0.94},
    xticklabels={},
    ]
      \addplot[dashed, samples=10, smooth,domain=0:1,black, name path=one] coordinates {(0.872941,0)(0.872941,8e-4)};
      \addplot[dashed, samples=10, smooth,domain=0:1,black, name path=two] coordinates {(0.901724,0)(0.901724,8e-4)};
      \addplot[dashed, samples=10, smooth,domain=0:1,black, name path=three] coordinates {(0.926152,0)(0.926152,8e-4)};
      \node[anchor=north west] (label1) at (axis cs: 0.86,7.0e-4) {\colorbox{white}{reference case}};
      \addplot[line width=0.75pt,mark=o,mark size=2.0pt,red] table[x expr=(\thisrowno{2}), y expr=(-\thisrowno{17} / \thisrowno{3}), col sep=space] {figs/Data/T453K_1atm_1e-0rrc/results_vary_k_FINAL.dat};
%
    \nextgroupplot
    [
    axis line style={-},
    xlabel={porosity [-]},
    xmin=0.86, xmax=0.94,
    ymin=1.6e-4, ymax=1.8e-4,
    x tick label style={
      /pgf/number format/.cd,
      fixed,
      fixed zerofill,
      precision=2,
      /tikz/.cd },
    y tick label style={
      /pgf/number format/.cd,
      fixed,
      fixed zerofill,
      precision=2,
      /tikz/.cd },
    xtick={0.86,0.87,...,0.94},
    ]
      \addplot[dashed, samples=10, smooth,domain=0:1,black, name path=one] coordinates {(0.872941,0)(0.872941,2e-4)};
      \addplot[dashed, samples=10, smooth,domain=0:1,black, name path=two] coordinates {(0.901724,0)(0.901724,2e-4)};
      \addplot[dashed, samples=10, smooth,domain=0:1,black, name path=three] coordinates {(0.926152,0)(0.926152,2e-4)};
      \node[anchor=south west] (label2) at (axis cs: 0.86,1.60e-4) {\colorbox{white}{active sites reduced to $0.01 \, \%$}};
      \addplot[line width=0.75pt,mark=o,mark size=2.0pt,red] table[x expr=(\thisrowno{2}), y expr=(-\thisrowno{17} / \thisrowno{3}), col sep=space] {figs/Data/T453K_1atm_1e-4rrc/results_vary_k_FINAL.dat};
%
  \end{groupplot}
  \node (y_label) at ($(group c1r1.north west)!0.5!(group c1r2.south west)$) [xshift=-1.0cm] {\rotatebox{90}{$\langle \eta \rangle / A_\text{s}$ [m]}};
\end{tikzpicture}
}
\caption{\zweizeilig To evaluate the surface utilization, we consider the effective conversion rate over the specific surface. In this study, the unit cell size is fixed, i.e., the actual surface area is directly proportional to the specific surface.}
\label{fig:surface_utilization}
\end{figure}
Further, the surface utilization rises with porosity, as shown in Figure~\ref{fig:surface_utilization}. For the specific surface, $A_\text{s}$, an analytical expression is given in~\cite{Smorygo2011}. For the same reason as above, the rise in surface utilization is less pronounced in the intermediate regime. 

\begin{figure}[!htb]
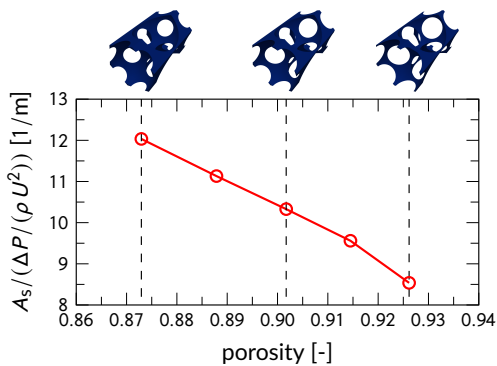

\centering
\tikzsetnextfilename{ObjectiveFunctionRRLR}
 \resizebox{\figWidth}{!}{%
\begin{tikzpicture}
  \begin{axis}
  [
    axis line style={-},
    xlabel={porosity [-]},
    ylabel={$A_\text{s} / ( \Delta P / (\rho \, U^2) )$ [$1/$m]},
    xlabel near ticks,
    ylabel near ticks,
    xmin=0.86, xmax=0.94,
    ymin=8, ymax=13,
    tick style={color=black},
    tick label style={font=\footnotesize},
    x tick label style={
      /pgf/number format/.cd,
      fixed,
      fixed zerofill,
      precision=2,
      /tikz/.cd },
    xtick={0.86,0.87,...,0.94},
    ytick={0,1,...,20},
    minor x tick num={1},
    minor y tick num={1},
    width=0.95\figWidth,
    height=0.60\figWidth,
  ]
  \addplot[line width=0.75pt,mark=o,mark size=2.0pt,red] table[x expr=(\thisrowno{2}), y expr=(\thisrowno{3} / (\thisrowno{8} * \thisrowno{11} * \thisrowno{11})), col sep=space] {figs/Data/T453K_1atm_1e-0rrc/results_vary_k_FINAL.dat};
%
  \addplot[dashed, samples=10, smooth,domain=0:1,black, name path=one] coordinates {(0.872941,4)(0.872941,20)};
  \addplot[dashed, samples=10, smooth,domain=0:1,black, name path=two] coordinates {(0.901724,4)(0.901724,20)};
  \addplot[dashed, samples=10, smooth,domain=0:1,black, name path=three] coordinates {(0.926152,4)(0.926152,20)};
  \end{axis}
  \node at (0.9,3.1){
    \includegraphics[width=0.15\figWidth]{figs/BasisCell_Examples/foam-1070.png}
  };
  \node at (2.65,3.1){
    \includegraphics[width=0.15\figWidth]{figs/BasisCell_Examples/foam-1090.png}
  };
  \node at (4.15,3.1){
    \includegraphics[width=0.15\figWidth]{figs/BasisCell_Examples/foam-1110.png}
  };
\end{tikzpicture}
}
\caption{\zweizeilig To estimate the suitability for the use as substrate in heterogeneous catalysis in the reaction rate limited regime, we consider the ratio between specific surface and dimensionless pressure drop.}
\label{fig:objectiveFunction_rrlr}
\end{figure}
Note that in the reaction rate limited regime, the concentrations of the reactants, and also $\langle \dot{n}_{r} \rangle$, are approximately constant throughout the system. Thus, the total reaction rate,
\begin{equation}
  \label{eq:rrlr}
  L \, \int \! \langle \dot{n}_{r} \rangle \, da \, = \, L \, \langle \dot{n}_{r} \rangle \, \int \! da \, ,
\end{equation}
where $da$ denotes the surface element, is directly proportional to the available surface area. For given inlet concentrations, from the particle balance one sees that also $\eta$ is proportional to the surface area. Since $\eta$ is typically small in this regime, $\langle \eta \rangle = -\ln(1 - \eta) \approx \eta$. Hence, to assess the performance index in the mass transfer limited regime, it is reasonable to regard the ratio between surface area and pressure drop. Figure~\ref{fig:objectiveFunction_rrlr} indicates that in the reaction rate limited regime, lower porosity values are beneficial, while according to Figure~\ref{fig:objectiveFunction} the opposite is true in the intermediate and mass transfer limited regime.

\section{Conclusion}

In this work, we have developed a technique for particle-based simulations of heterogeneous catalysis in open-cell foam structures, consisting of four main components. The foam geometry is modeled as an inverse sphere packing using CSG~\cite{Maentylae1987, Strobl2017}. To eliminate any anisotropy at the complex boundaries, an isotropic variant of SRD is employed to simulate the gas dynamics~\cite{Muehlbauer2017}. The inlet and outlet are simulated with semi-periodic boundary conditions~\cite{Muehlbauer2018}. Finally, the heterocatalytic reaction in the washcoat is decoupled from the particle simulation by introducing an effective reaction model.

Assuming the low temperature water-gas shift reaction, we have validated the developed simulation technique based on experimental data. We have found the relation between  Hagen and Reynolds number to agree with experimental findings, especially at low Reynolds numbers~\cite{Dukhan2006}. Moreover, within the Reynolds number range covered in the experiments, the Sherwood number in the simulations qualitatively agrees with experimental data~\cite{Giani2005}. Interestingly, for smaller Reynolds numbers, we find a qualitatively different behavior of the Sherwood number, indicating the transition between two different flow regimes at $\text{Re}_\text{\Giani} \approx 10$.

With this simulation setup, we are able to assess reactive flows depending on the reaction parameters in the washcoat. As a starting point, we have assumed typical parameters from literature~\cite{Ayastuy2005,Novak2012}. We have found that, even in the small Reynolds number regime, the number of active sites may be reduced by a factor of $100$ with respect to this reference configuration without a notable change in conversion efficiency.

Further, we have examined the performance index and surface utilization for different values of porosity and \Thiele{}. For the considered setup, the surface utilization improves as the porosity is increased, both in the mass transfer limited and in the intermediate \Thiele{} regime. In accordance with literature~\cite{Lucci2014}, we have found that the performance index also rises with increasing porosity in the mass transfer limited regime. The same is true in the intermediate \Thiele{} regime, however, as for the surface utilization, the effect is less pronounced. Conversely, in the reaction rate limited regime, the surface utilization is expected to stay roughly constant, while the performance index, consequently, decreases as porosity is increased.


\section*{Acknowledgements}

We thank Michael Blank for sharing his expertise in chemical reaction engineering.
We also thank the German Research Foundation (DFG) for funding through the Cluster of Excellence ``Engineering of Advanced Materials'', ZISC, and, FPS.



\printendnotes

\bibliographystyle{aichej}
\bibliography{CatalyticFoam}


\end{document}